\documentclass[pra,superscriptaddress,twocolumn,aps,longbibliography]{revtex4-1}
\usepackage{epsfig}             
\usepackage{epstopdf}           
\usepackage{hyperref}           
\usepackage{color}
\usepackage{colortbl}
\usepackage{graphicx}
\usepackage{amssymb,amsmath}
\usepackage{subfig}
\usepackage{caption}
\usepackage{enumerate}
\usepackage{gensymb}
\usepackage{url}
\usepackage{floatrow}

\begin{document}

\title[Theoretical Atto-nano Physics]{Theoretical Atto-nano Physics}

\author{Marcelo F. Ciappina}
\affiliation{Institute of Physics of the ASCR, ELI-Beamlines, Na Slovance 2, 182 21 Prague, Czech Republic}

\author{Maciej Lewenstein}
\affiliation{ICFO - Institut de Ciencies Fotoniques, The Barcelona Institute of Science and Technology, Av. Carl Friedrich Gauss 3, 08860 Castelldefels (Barcelona), Spain}
\affiliation{ICREA - Instituci\'{o} Catalana de Recerca i Estudis Avan\c{c}ats, Lluis Companys 23, 08010 Barcelona, Spain}

\date{\today}

\begin{abstract}
Two emerging areas of research, attosecond and nanoscale physics, have recently started to merge. Attosecond physics deals with phenomena occurring when ultrashort laser pulses,
with duration on the femto- and sub-femtosecond time scales, interact with atoms, molecules
or solids. The laser-induced electron dynamics occurs natively on a timescale down to a few hundred or even tens of attoseconds (1 attosecond=1 as=10$^{-18}$ s), which is of the order of the optical field cycle. For comparison, the revolution of an electron on a $1s$ orbital of a hydrogen atom is $\sim152$ as. On the other hand, the second topic involves the manipulation and engineering of mesoscopic systems, such as solids, metals and dielectrics, with nanometric precision. Although nano-engineering is a vast and well-established research field on its own, the combination with intense laser physics is relatively recent. We present a comprehensive theoretical overview of the tools to tackle and understand the physics that takes place when short and intense laser pulses interact with nanosystems, such as metallic and dielectric nanostructures. In particular we elucidate how the spatially inhomogeneous laser induced fields at a nanometer scale modify the laser-driven electron dynamics. Consequently, this has important impact on pivotal processes such as above-threshold ionization and high-order harmonic generation. The deep understanding of the coupled dynamics between these spatially inhomogeneous fields and matter configures a promising way to new avenues of research and applications. Thanks to the maturity that attosecond physics has reached, together with the tremendous advance in material engineering and manipulation techniques, the age of atto-nano physics has begun, but it is still in an incipient stage. 

\smallskip

This document is the unedited Author's version of a Submitted Work that was subsequently accepted for publication in the \textit{21st Century Nanoscience – A Handbook: Nanophysics Sourcebook (Volume One)}, copyright \copyright\, CRC Press, Taylor \& Francis Group after peer review. The final edited and published work is available at 
\url{https://doi.org/10.1201/9780367333003}
\end{abstract}

\maketitle

\section{Introduction}\label{intro}

This chapter deals with an embryonic field of atomic, molecular, and optical physics: atto-nanophysics. It is an area that combines the traditional and already very mature attosecond physics with the equally well developed nanophysics. We start our contribution by just presenting the general motivations and description of this new area, restricting ourselves to very basic and well known concepts and references. 

Attosecond physics has traditionally focused on atomic and small molecular targets~\cite{Scrinzi06,Krausz09}.  For such systems the electron, once it is ionized by the electric field of the laser pulse, moves in a region that is small compared to the wavelength of the driving laser. Hence,  the spatial dependence of the laser field can be safely neglected.  In the presence of such so-called 'spatially homogeneous' laser fields the time-dependent processes occurring on the attosecond time scale have been extensively investigated~\cite{Krausz01,baltuska_attosecond_2003}.  This subject has came to age based upon well-established theoretical developments and the understanding of numerous nonlinear phenomena (cf.~\cite{Salieres-adv,Joachain2, MaciejChapter}), as well as the tremendous advances in experimental laser techniques.  For instance, nowadays experimental measurements with attosecond precision are routinely performed in several facilities around the world. 

Simultaneously, bulk matter samples have been downscaled in size to nanometric dimensions, opening the door to study light-matter interaction in a completely new arena. When a strong laser interacts, for instance, with a metallic structure, it can couple with the plasmon modes inducing the ones corresponding to collective oscillations of free charges (electrons). These free charges, driven by the field, generate 'nanospots', of few nanometers size, of highly enhanced near-fields, which exhibit unique temporal and spatial properties. The near-fields, in turn, induce appreciable changes in the local field strength at a scale of the order of tenths of nanometers, and in this way they play an important role modifying the field-induced electron dynamics. In other words, in this regime, the spatial scale on which  the electron dynamics takes place is of the same order as the field variations. Moreover,  the near-fields change on a sub-cycle timescale as the free charges respond almost instantaneously to the driving laser. Consequently, we face an unprecedented scenario: the possibility to study and manipulate strong-field induced processes by rapidly changing fields, which are not spatially homogeneous. In the following subsections we present a description of these strong field processes, joint with the theoretical tools particularly modified to tackle them.  The emergent field of attosecond physics at the nanoscale marries very fast attosecond processes (1 as = $10^{-18}$ s), with very small nanometric spatial scales (1 nm = $10^{-9}$ m), bringing a unique and sometimes unexpected perspective on important underlying strong field phenomena.


\subsection{Strong field phenomena driven by spatially homogeneous fields}

A particular way of initiating electronic dynamics in atoms, molecules and recently solid materials, is to expose these systems to an intense and coherent electromagnetic radiation.  A variety of widely studied and important phenomena, which we simply list and shortly describe here, result from this interaction. In order to first put the relevant laser parameters into context, it is useful to compare them with an atomic reference. In the present framework, laser fields are considered intense when their strength is not much smaller or even comparable to the Coulomb field experienced by an atomic electron. The Coulomb field in an hydrogen atom is approximately $5\times10^9$ V cm$^{-1}$ ($\approx 514$ V nm$^{-1}$), corresponding to an equivalent intensity of $3.51\times10^{16}$ W cm$^{-2}$ -- this last value indeed defines the atomic unit of intensity. With regard to time scales, we note that in the Bohr model of hydrogen atom, the electron takes about 150 as to orbit around the proton, defining thus the characteristic time for electron dynamics inside atoms and molecules~\cite{Krausz07}.  Finally, the relevant laser sources are typically in the near-infrared (NIR) regime, and hence laser frequencies are much below the ionisation atomic or molecular potential.  In particular, an 800 nm source corresponds to a photon energy of 1.55 eV ($0.057$ au), which is much below the ionisation potential of hydrogen, given by $1/2$ au (13.6 eV).  At the same time, laser intensities are in the $10^{13} - 10^{15}$ W cm$^{-2}$ range: high enough to ionize a noticeable fraction of the sample, but low enough to avoid space charge effects and full depletion of the ground state.

While the physics of interactions of atoms and molecules with intense laser pulses is quite complex, much can be learnt using theoretical tools developed over the past decades, starting with the original work by Keldysh in the 1960's~\cite{Keldysh,PPT1966,Reiss,ADK1986,FaisalBook}.  According to the Keldysh theory, an electron can be freed from an atomic or molecular core either via tunnel or multiphoton ionization. These two regimes are characterized by the Keldysh parameter:

\begin{equation}
\gamma=\omega_0\frac{\sqrt{2I_p}}{E_0}=\sqrt{\frac{I_p}{2U_p}},
\end{equation}
where $I_p$ is the ionization potential, $U_p$ is the ponderomotive energy, defined as $U_p=E_0^2/ 4 \omega_0^2$ where $E_0$ is the peak laser electric field and $\omega_0$ the laser carrier frequency. The adiabatic tunnelling regime is then characterized by $\gamma\ll 1$, whereas the multiphoton ionization regime by $\gamma \gg 1$. In the multiphoton regime ionisation rates scale as laser intensity $I^N$, where $N$ is the order of the process, i.e.~the number of photon necessary to overpass the ionization potential. 

Many experiments take place in an intermediate or {\it cross-over} region, defined by $\gamma\sim 1$~\cite{AttoTunnelTime}.  Another way to interpret $\gamma$ is to note that $\gamma=\tau_T/\tau_L$, where $\tau_T$ is the Keldysh time (defined as $\tau_T = \frac{\sqrt{2I_p}}{E_0}$) and $\tau_L$ is the laser period.  Hence  $\gamma$ serves as a measure of non-adiabaticity by comparing the response time of the electron wavefunction to the period of the laser field.  

When laser intensities approach $10^{13}\sim10^{14}$ W cm$^{-2}$, the usual perturbative scaling  observed in the multiphoton regime ($\gamma\gg 1$) does not hold, and the emission process becomes dominated by tunnelling ($\gamma<1$). In this regime a strong laser field bends the binding potential of the atom creating a penetrable potential barrier. The ionization process is governed thus by electrons tunnelling through this potential barrier, and subsequently interacting "classically" with the strong laser field far from the parent ion~\cite{corkum93, schafer93,Lewenstein94}.  

This concept of tunnel ionization underpins many important theoretical advances, which have received crystal clear experimental confirmation with the development of intense ultra-short lasers and attosecond sources over the past two decades.  On a fundamental level, theoretical and experimental progress opened the door to the study of basic atomic and molecular processes on the attosecond time scale.  On a practical level, this led to the development of attosecond high frequency extreme ultraviolet (XUV) and X-ray sources, which promise many important applications, such  fine control of atomic and molecular reactions among others. The very fact that we deal here with sources that produce pulses of attosecond duration is remarkable. Attosecond XUV pulses allow, in principle,  to capture all processes underlying structural dynamics and chemical reactions, including electronic motion coupled to nuclear dynamics. They allow also to address basic unresolved and controversial questions in quantum mechanics, such as, for instance,  the duration of the strong field ionization process or the tunnelling time~\cite{AttoTunnelTime, pazourek15}.

Among the variety of phenomena which take place when atomic systems are driven by coherent and intense electromagnetic radiation, the most notable examples are high-order harmonic generation (HHG), above-threshold ionization (ATI) and multiple sequential or non-sequential ionization.  All these processes present similarities and differences, which we detail briefly below~\cite{Joachain1, Joachain2, MaciejChapter}
 
HHG takes place whenever an atom or molecule interacts with an intense laser field of frequency $\omega_0$, producing radiation of higher multiples of the fundamental frequency $K\omega_0$, where in the simplest case of rotationally symmetric target $K$ is an odd integer. HHG spectra present very distinct characteristics: there is a sharp decline in conversion efficiency followed by a plateau in which the harmonic intensity hardly varies with the harmonic order $K$, and eventually an abrupt cutoff.  For an inversion symmetric medium (such as all atoms and some molecules), only odd harmonics of the driving field have been observed because of dipole selection rules and the central symmetric character of the potential formed by the laser pulse and the atomic field. The discovery of this plateau region in HHG has made the generation of coherent XUV radiation using table-top lasers feasible. The above mentioned features characterize a highly nonlinear process~\cite{AnneHHG}. Furthermore, HHG spectroscopy (i.e.~the measurement and interpretation of the HHG emission from a sample) has been widely applied to studying the ultrafast dynamics of molecules interacting with strong laser fields (see, e.g.~\cite{marangos2016Review}).

Conceptually, HHG is easily understood using the three-step model~\cite{kulander,corkum93,Lewenstein94,muller,kuchiev1987}: (i) tunnel ionization due to the intense and low frequency laser field; (ii) acceleration of the free electron by the laser electric field, and (iii) re-collision with the parent atom or molecular ion. The kinetic energy gained by the electron in its travel, under the presence of the laser oscillatory electric field, is converted into a high energy photon and can be easily calculated starting from semiclassical assumptions. 

HHG has received special attention because it configures the workhorse for the creation of attosecond pulses and, simultaneously, it exemplifies a special challenge from a theoretical point of view due to the complex intertwining between the Coulomb and external laser fields. Additionally, HHG is a promising way to provide a coherent table-top sized short wavelength light sources in the XUV and soft x-ray regions of the spectrum. Nonlinear atom-electron dynamics triggered by focusing intense laser pulses onto noble gases generates broadband high photons whose energy reaches the soft X-ray region. This nonlinear phenomenon requires laser intensities in the range of $10^{14}$ W cm$^{-2}$, routinely available from Ti:sapphire femtosecond laser amplifiers~\cite{Krausz00}.

Another widely studied phenomenon is ATI.  In fact, and from an historical viewpoint, it was the first one to be considered as a strong nonperturbative laser-matter interaction process~\cite{VanderWiel,AgostiniATI}.  Conceptually, ATI is similar to HHG, except the electron does not recombine with the parent atom in the step (iii), but rather it is accelerated away by the laser field, eventually being registered at the detector.   Hence, ATI is a much more likely process than HHG, although the latter has opened a venue for a larger set of applications and technological developments. Nevertheless, ATI is an essential tool for laser pulse characterization, in particular in the few-cycle pulses regime.  Unlike in HHG, where macroscopic effects, such as phase matching, often have to be incorporated to reliably reproduce the experiment, single atom simulations are generally enough for ATI modeling. 

In an ordinary ATI experiment, the energy and/or angular distribution of photoelectrons is measured. The ATI spectrum in energy presents a series of peaks given by the formula $E_p=(m+s)\omega_0-I_p$, where $m$ is the minimum number of laser photons needed to exceed the atomic binding energy $I_p$ and $s$ is commonly called the number of `above-threshold' photons carried by the electron. This picture changes dramatically when few-cycle pulses are used to drive the media and the ATI energy spectra become much richer structurally speaking~\cite{Milosevic06}.

In this case, we can clearly distinguish two different regions, corresponding to the direct and rescattered electrons, respectively.  The low energy region, given by $E_k\lesssim2 U_p$, corresponds to direct electrons or electrons which never come back to the vicinity of the parent atom. On the other hand, the high energy part of the ATI spectrum $2 U_p\lesssim E_k\lesssim 10 U_p$ is dominated by the rescattered electrons, i.e.~the electrons that reach the detector after being rescattered by the remaining ion-core~\cite{Paulusplateau}. The latter are strongly influenced by the absolute phase of a few-cycle pulse and as a consequence they are used routinely for laser pulse characterization~\cite{paulus_measurement_2003}. These two energy limits for both the direct and rescattered electrons, i.e.~$2U_p$ and $10 U_p$ can be easily obtained invoking purely classical arguments~\cite{Becker02Chapter,Milosevic06,salieres2001}.

Most of the ATI and HHG experiments use as an interacting media multielectronic atoms and molecules, and recently condensed and bulk matter.  Nevertheless, one often assumes that only one valence electron is active and hence determines all the significant features of the strong field laser-matter interaction.  The first observations of two-electron effects in ionization by strong laser pulses go back to the famous Anne L'Huillier's `knee'~\cite{Anne-knee}. This paper and later the influential Paul Corkum's work~\cite{corkum93} stimulated the discussion about sequential versus non-sequential ionization, and about a specific mechanism of the latter (shake-off, rescattering, etc.). In the last twenty years, and more recently as well, there has been a growing interest in electron correlations, both in single- and multi-electron ionization regimes, corresponding to lower and higher intensities, respectively (cf. ~\cite{Walker94,olgaHHG,Corkumcollective}).

One prominent example where electron correlation plays an instrumental role is the so-called non-sequential double ionization (NSDI)~\cite{Walker94}. It stands in contrast to sequential double (or multiple) ionization, i.e.~when the process undergoes a sequence of single ionization events, with no correlation between them. NSDI has attracted considerable interest, since it gives direct experimental access to electron-electron correlation -- something that is famously difficult to analyse both analytically and numerically (for a recent review see, e.g.~\cite{bergues15}).  

\subsection{Introduction to atto-nano physics}



The interaction of ultra-short strong laser pulses with extended systems has recently received much attention and led to an advance in our understanding of the attosecond to few-femtosecond electronic and nuclear dynamics. For instance, the interaction of clusters with strong ultrafast laser fields has long been known to lead to the formation of nanoplasmas in which there is a high degree of charge localisation  and ultrafast dynamics, with the emission of energetic (multiple keV) electrons  and highly charged -up to Xe$^{40+}$- ions with high energy (MeV scale)~\cite{Shao1996,DitmireNature,Ditmire1997,Smith1998,Tisch1997}. Most recently use of short pulses ($\sim10$ fs)  has succeeded in isolating the electron dynamics from the longer timescale  ion dynamics (which are essentially frozen) revealing a higher degree of fragmentation  anisotropy in both electrons and ions compared to the isotropic distributions found from longer pulses ($\sim100$ fs)~\cite{Skopalova2010}.

Likewise, interactions of intense lasers with nano-particles, such as micron scale liquid droplets,  leads to hot plasma formation. An important role is found for enhanced local fields on the surface of  these droplets driving this interaction via ``field hotspots''~\cite{Symes2004,Gumbrell2001,Donnelly2001,Mountford1998,Sumeruk2007Plasmas,Sumeruk2007}.

Furthermore, studies of driving bound and free charges in larger molecules, e.g.~collective electron dynamics in fullerenes~\cite{li_etal2015}, and in graphene-like structures~\cite{baum2015}, proton migration in hydrocarbon molecules~\cite{kuebel2015}, and charge migration in proteins~\cite{calegari2012,calegari2014} could be included in this group. In turn, laser-driven broad-band electron wavepackets have been used for static and dynamic diffraction imaging of molecules~\cite{blaga2012,xu2014,mick2015}, obtaining structural information with sub-nanometer resolution.

Tailored ultra-short and intense fields have also been used to drive electron dynamics and electron or photon emission from (nanostructured) solids (for a recent compilation see e.g.~\cite{Hommelhoff15}). The progress seen in recent years has been largely driven by advances in experimental and engineering techniques (both in laser technology and in nanofabrication). Among the remarkable achievements in just the latest years are the demonstration of driving electron currents and switching the conductivity of dielectrics with ultrashort pulses~\cite{schiffrin2013, schultze2013controlling}, controlling the light-induced electron emission from nanoparticles~\cite{Suessmann15,Zherebtsov11} and nanotips~\cite{Kruger11,Herink12,Piglosiewicz2014}, and the sub-cycle driven photon emission from solids~\cite{Ghimire2011,Trang2015,Huber2014,Vampa2015}. Furthermore, the intrinsic electron propagation and photoemission processes have been investigated on their natural, attosecond timescales~\cite{schultze2010delay,neppl2012attosecond,cavalieri2007attosecond,locher2015,okell_temporal_2015}.

\begin{figure}
	\centering	
	\includegraphics[width=.85\columnwidth]{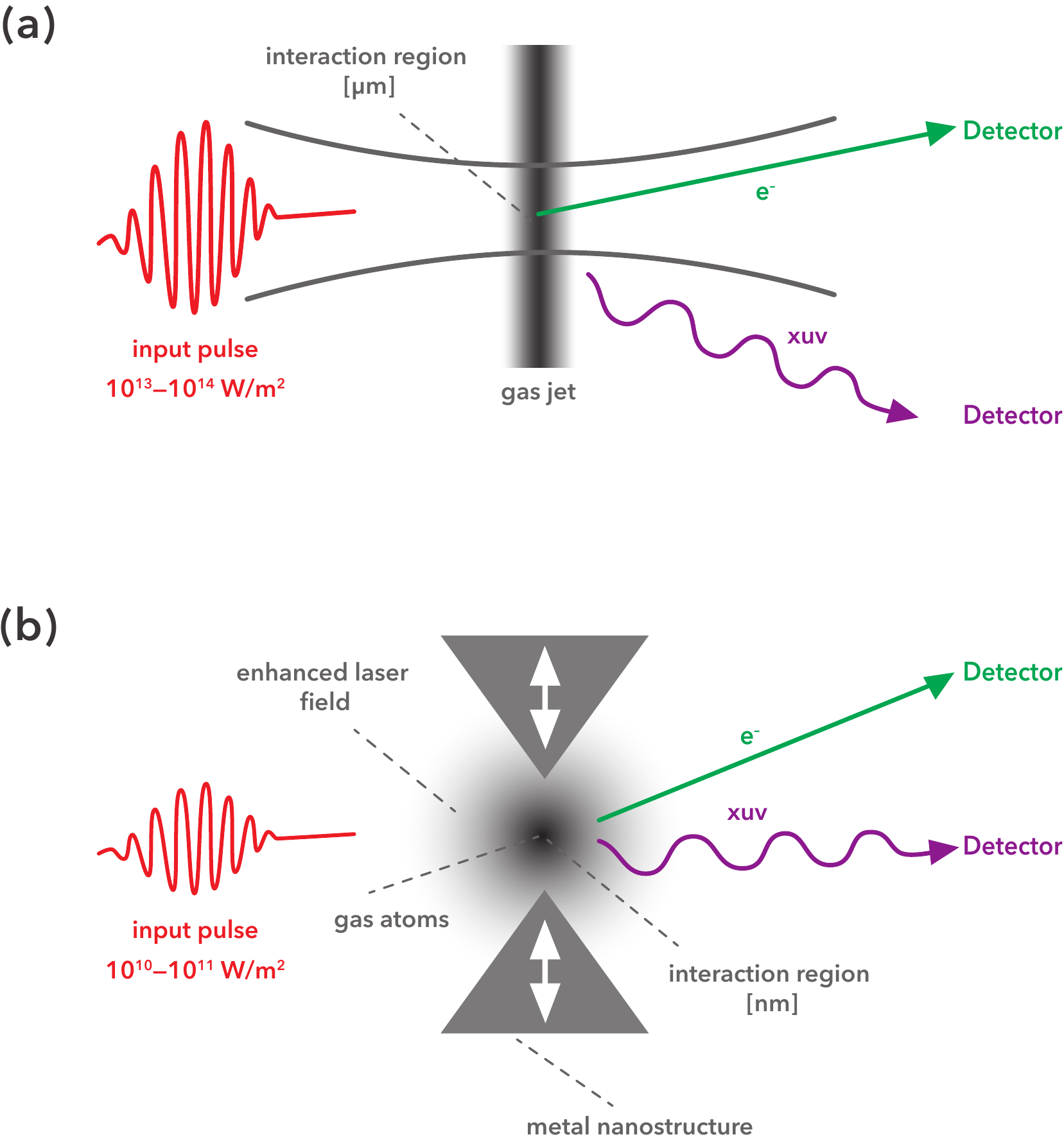}
	\caption{Sketch of conventional (a) and plasmonic-enhanced (b) strong field processes.}
	\label{Intro-fig:sketch}
\end{figure}

A key feature of light-nanostructure interaction is the enhancement (amplification) of the electric near-field by several orders of magnitude, and its local confinement on a sub-wavelength scale~\cite{stockmanreview}.  From a theoretical viewpoint, this field localisation presents a unique challenge:  we have at our disposal strong fields that  change on a comparable spatial scale of the oscillatory electron dynamics that are initiated by those same fields.  As will be shown throughout this contribution, this singular property entails profound consequences in the underlying physics of the conventional strong field phenomena.  In particular, it defies one of the main assumptions that modelling of strong-field interactions is based upon: the spatial homogeneity of laser fields in the volume of the electronic dynamics under scrutiny. 

Interestingly, an exponential growing attention in strong field phenomena induced
by plasmonic-enhanced fields was triggered by the questionable work of Kim et al.~\cite{Kim08}.
These authors claimed to have been observed efficient HHG from bow-tie metallic nanostructures.
Although the interpretation of the outcomes was incorrect, this paper definitively 
stimulated a constant interest in the plasmonic-enhanced HHG and ATI~\cite{Sivis13,Park11,Kovacev13NJP,Sivis12A,Kim12Reply,Park13}.

 Within the conventional assumption, both the laser electric field, $E(\mathbf{r},t)$, and the corresponding vector potential, $A(\mathbf{r},t)$, are spatially homogeneous in the region where the electron moves and only their time dependence is considered, i.e.~$E(\mathbf{r},t)=E(t)$ and $A(\mathbf{r},t)=A(t)$. This is a authentic assumption considering the usual electron excursion (estimated classically using $\alpha=E_0/\omega_0^2$) is bounded roughly by a few nanometers in the NIR, for typical laser intensities, and several tens of nanometers for mid-infrared (MIR) sources (note that $\alpha\propto \lambda_0^2$, where $\lambda_0$ is the wavelength of the driving laser and $E_0=\sqrt{I}$, where $I$ is the laser intensity)~\cite{Krausz00}. Hence, electron excursion is very small relative to the spatial variation of the field in the absence of local (or nanoplasmonic) field enhancement (see Fig.~\ref{Intro-fig:sketch}(a)). On the contrary, the fields generated using surface plasmons are spatially dependent on a nanometric region (cf. Fig.~\ref{Intro-fig:sketch}(b)).  As a consequence, all the standard theoretical tools in the strong field ionization toolbox (ranging from purely classical to frequently used semiclassical and complete quantum mechanical approaches) have to be re-pondered.  In this contribution, we will therefore concentrate our efforts on how the most important and basic processes in strong field physics, such as HHG and ATI, are modified in this new setting of strong field ultrafast phenomena on a nano-scale. Additionally, we discuss how the conventional theoretical tools have to be modified in order to be suitable for this new scenario.  Note that the strong field phenomena driven by plasmonic fields could be treated theoretically within a particular flavour of a non-dipole approximation, but neglecting completely magnetic effects. 
 
\section{Theoretical approaches}

In the next subsections we describe the theoretical approaches we have developed to tackle strong field processes driven by spatially inhomogeneous laser fields. We put particular emphasis on HHG and ATI.

\subsection{Quantum approaches}

The dynamics of a single active atomic electron in a strong laser field takes place along the polarization direction of the field, when linearly polarized laser pulses drive the system. It is then perfectly legitimate to model the HHG and ATI in a 1D spatial dimension by solving the time-dependent Schr\"odinger equation (1D-TDSE)~\cite{Marcelo12A}:
\begin{eqnarray}
\label{tdse1d}
\mathrm{i} \frac{\partial \Psi(x,t)}{\partial t}&=&\mathcal{H}(t)\Psi(x,t) \\
&=&\left[-\frac{1}{2}\frac{\partial^{2}}{\partial x^{2}}+V_{\rm{a}}(x)+V_{\rm{l}}(x,t)\right]\Psi(x,t), \nonumber
\end{eqnarray}
where in order to model an atom in 1D, it is common to use soft core potentials, which are of the form:
\begin{equation}
V_{\rm{a}}(x)=-\frac{1}{\sqrt{x^2+b^2}},
\end{equation} 
where the parameter $b$ allows us to modify the ionization potential $I_p$ of the ground state, fixing it as close as possible to the value of the atom under consideration.
We consider the field to be linearly polarized along the $x$-axis and modify the interaction term $V_{\rm{l}}(x,t)$ in order to treat spatially
nonhomogeneous fields, while maintaining the dipole character.
Consequently we write
\begin{eqnarray}  
\label{vlaser1d}
V_{\rm{l}}(x,t)&=&-E(x,t)\,x
\end{eqnarray}
where $E(x,t)$ is the laser electric field defined as
\begin{equation}  
\label{electric}
E(x,t)=E_0\,f(t)\, (1+\varepsilon h(x))\,\sin(\omega_0 t+\phi).
\end{equation}
In Eq.~(\ref{electric}), $E_0$, $\omega_0$ and $\phi$ are the peak
amplitude, the frequency of the laser pulse and the carrier-envelope phase (CEP), respectively. We refer to sin(cos)-like laser pulses where $\phi=0$ ($\phi=\pi/2$). The pulse envelope is given by $f(t)$ and $\varepsilon$ is a small parameter that characterizes the inhomogeneity strength. The function $h(x)$
represents the functional form of the spatial nonhomogeneous field and, in
principle, could take any form and be supported by the numerical algorithm (for details see e.g.~\cite{Marcelo12A,Marcelo12OE}). Most of the approaches use the simplest form for $h(x)$, i.e.~the linear term: $h(x)=x$. This choice is motivated by previous investigations~\cite{Husakou11A,Marcelo12A,Yavuz12}, but nothing prevents to use more general functional forms for $h(x)$~\cite{EnriqueJOpt}. 

The actual spatial dependence of the enhanced near-field in the surrounding of a metal nanostructure can be obtained by solving the Maxwell equations incorporating both the geometry and material properties of the nanosystem under study and the input laser pulse characteristics (see e.g.~\cite{Marcelo12OE}). The electric field retrieved numerically is then approximated using a power series  $h(x) =\sum_{i=1}^{N}b_{i}x^{i}$, where the coefficients $b_i$ are obtained by fitting the real electric field that results from a finite element simulation. Furthermore, in the region relevant for the strong field physics and electron dynamics and in the range of the parameters we are considering, the electric field can be indeed approximated by its linear dependence.

The 1D-TDSE can be solved numerically by using the Crank-Nicolson scheme in order to obtain the time propagated electronic wavefunction $\Psi(x,t)$. Once $\Psi(x,t)$ is found, we can compute the harmonic spectrum $D(\omega)$ by Fourier transforming the dipole acceleration $a(t)$ of the active electron.  That is, 
\begin{equation}
D(\omega)=\left| \frac{1}{T_p}\frac{1}{\omega^2}\int_{-\infty}^{\infty}dt e^{-i\omega t} a(t)\right|^2,
\end{equation}
where $T_p$ is the total duration of the laser pulse. $a(t)$ can be obtained by using the commutator relation
\begin{equation}
a(t)=\frac{d^2\langle x \rangle}{dt^2}=-\langle \Psi(t)| \left[\mathcal{H}(t),\left[ \mathcal{H}(t),x\right]\right]|\Psi(t)\rangle,
\end{equation}
where $\mathcal{H}(t)$ is the Hamiltonian specified in Eq.~(\ref{tdse1d}).

One of the main advantages of the 1D-TDSE is that we are able to include any functional form for the spatial variation of the plasmonic field. For instance, we have implemented linear~\cite{Marcelo12A} and real (parabolic) plasmonic fields~\cite{Marcelo12OE}, as well as near-fields with exponential decay (evanescent fields)~\cite{Marcelo13AR} and gaussian-like bounded spatially fields~\cite{EnriqueJOpt}. 


In order to calculate ATI-related observables, we use the same one-dimensional time-dependent Schr\"odinger equation (1D-TDSE) employed for the computation of HHG [Eq.~(\ref{tdse1d})]. In 1D we are only able to compute the so-called energy-resolved photoelectron spectra $P(E)$, i.e.~a quantity proportional to the probability to find electrons with a particular energy $E$. In order to do so, we use the window function technique developed by Schafer~\cite{schaferwop1,schaferwop}. 
This tool has been widely used, both to calculate angle-resolved and
energy-resolved photoelectron spectra~\cite{schaferwop2} and it represents a
step forward with respect to the usual projection methods.

An extension of the above described approach is to solve the TDSE in its full dimensionality and to include in the laser-electron potential the spatial variation of the laser electric field. For only one active electron we need to deal with 3 spatial dimensions and, due to the cylindrical symmetry of the problem, we are able to separate the electronic wavefunction in spherical harmonics, $Y_l^m$ and consider only terms with $m=0$ (see below). 

In particular, the 3D-TDSE in the length gauge can be written:  
\begin{eqnarray}
\label{tdse3d}
\nonumber
i\frac{\partial \Psi({\bf{r}},t)}{\partial t}&=&H\Psi({\bf{r}},t)\\
&=&\left [-\frac{\nabla^{2}}{2}+V_{SAE}({\bf{r}})+V_l({{\bf{r}},t})\right ]\Psi({\bf{r}},t),
\end{eqnarray}
where $V_{SAE}({\bf{r}})$ is the atomic potential in the single active electron (SAE) approximation and $V_l({{\bf{r}},t})$ the laser-electron coupling (see below).
The time-dependent electronic wave function $\Psi({\bf{r}},t)$, can be expanded in terms of spherical harmonics as:
\begin{eqnarray}
\label{spherical}
\nonumber
\Psi({\bf{r}},t)&=&\Psi({r,\theta, \phi},t)\\
&\approx&\sum_{l=0}^{L-1}\sum_{m=-l}^{l}\frac{\Phi_{lm}(r,t)}{r}Y_{l}^{m}(\theta,\phi)
\end{eqnarray}
where the number of partial waves depends on each specific case. Here, in order to assure the numerical convergence, we have used up to $L\approx250$ in the most extreme case ($I\sim 5\times10^{14}$ W/cm$^{2}$). In addition, due to the fact that the plasmonic field is linearly polarized, the magnetic quantum number is conserved and consequently in the following we can consider only $m=0$ in Eq.~(\ref{spherical}). This property considerably reduces the complexity of the problem. In here, we consider $z$ as a polarization axis and we take into account that the spatial variation of the electric field is linear with respect to the position. As a result, the coupling
$V_{l}(\mathbf{r},t)$ between the atomic electron and the
electromagnetic radiation reads
\begin{equation}
\label{vlaserati} V_{l}(\mathbf{r},t)=\int ^\mathbf{r}
d\mathbf{r'}\cdot\mathbf{E}(\mathbf{r'},t)=E_0z(1+\varepsilon
z)f(t)\sin(\omega_0 t+\phi)
\end{equation}
where $E_0$,  $\omega_0$ and $\phi$ are the laser electric field
amplitude, the central frequency and the CEP, respectively. As in previous investigations, the parameter $\varepsilon$ defines
the `strength'  of the inhomogeneity and has units of inverse
length (see also~\cite{Husakou11A,Yavuz12,Marcelo12A}). For modeling
short laser pulses in Eq.~(\ref{vlaserati}), we use a sin-squared
envelope $f(t)$ of the form $
f(t)=\sin^{2}\left(\frac{\omega_0 t}{2 n_p}\right)$,
where $n_p$ is the total number of optical cycles. As a result,
the total duration of the laser pulse will be $T_p=n_p \tau_L$ where
$\tau_L=2\pi/\omega_0$ is the laser period. We focus our analysis on a hydrogen atom, i.e.~$V_{SAE}({\bf{r}})=-1/r$ in Eq.~(\ref{tdse3d}), and we also assume that before
switch on the laser ($t=-\infty$) the target atom is
in its ground state ($1s$), whose analytic form can be found in a
standard textbook. Within the SAE
approximation, however, our numerical scheme is tunable to treat
any complex atom by choosing the adequate effective (Hartree-Fock)
potential $V_{SAE}({\bf{r}})$, and finding the ground state by the means of numerical
diagonalization.

Next, we will show how the inhomogeneity modifies the equations which model 
the laser-electron coupling. Inserting Eq.~(\ref{spherical}) into Eq.~(\ref{tdse3d}) and considering that,
\begin{equation}
\cos \theta Y_{l}^{0}=c_{l-1}Y_{l-1}^{0}+c_{l}Y_{l+1}^{0}
\end{equation}
and
\begin{equation}
\label{square}
\cos^{2} \theta Y_{l}^{0}=c_{l-2}c_{l-1}Y_{l-1}^{0}+(c_{l-1}^{2}+c_{l}^{2})Y_{l}^{0}+c_{l}c_{l+1}Y_{l+2}^{0},
\end{equation}
where
\begin{equation}
c_{l}=\sqrt{\frac{(l+1)^2}{(2l+1)(2l+3)}},
\end{equation}
we obtain a set of coupled differential equations for each of the radial functions $\Phi_{l}(r,t)$:
\begin{eqnarray}
\label{diag}
i\frac{\partial\Phi_{l}}{\partial{t}}=\left [-\frac{1}{2}\frac{\partial^{2}}{\partial r^{2}}+\frac{l(l+1)}{2r^2}-\frac{1}{2} \right ]\Phi_{l}\nonumber\\
+\varepsilon r^{2}E(t)\left(c_{l}^{2}+c_{l-1}^{2}\right)\Phi_{l}\nonumber\\
+r E(t)\left(c_{l-1}\Phi_{l-1}+c_{l}\Phi_{l+1}\right)\nonumber\\
+\varepsilon r^{2}E(t)\left(c_{l-2}c_{l-1}\Phi_{l-2}+c_{l}c_{l+1}\Phi_{l+2}\right).
\end{eqnarray}
Equation (\ref{diag}) is solved using the Crank-Nicolson algorithm considering the additional term, i.e.~Eq.~(\ref{square}) due to the spatial inhomogeneity. As can be observed, the degree of complexity will increase substantially when a more complex functional form for the spatial inhomogeneous laser electric field is used. For instance, the incorporation of only a linear term couples the angular momenta $l,l\pm1,l\pm2$, instead of $l,l\pm1$, as in the case of conventional (spatial homogeneous) laser fields.

As was already mentioned, typically several hundreds of angular momenta $l$ should to be considered and we could recognize the time evolution of each of them as a 1D problem. We use a Crank-Nicolson method implemented on a splitting of the time-evolution operator that preserves the norm of the wave function for the time propagation, similar to the 1D-TDSE case. The harmonic spectrum $D(\omega)$ is then computed in the same was as in the 1D case, but now using the 3D electronic wavefunctions $\Psi({\bf{r}},t)$.

We have also made studies on helium because a majority of experiments in HHG are carried out in noble gases. Nonetheless, other atoms could be easily implemented by choosing the appropriate atomic model potential $V_{SAE}({\bf{r}})$. After time propagation of the electronic wavefunction, the HHG spectra can be computed in an analogous way as in the case of the 1D-TDSE. Due to the complexity of the problem, only simulations with nonhomogeneous fields with linear spatial variations along the laser polarization in the 3D-TDSE have been studied. This, however, is enough to confirm that even a small spatial inhomogeneity significantly modifies the HHG spectra (for details see~\cite{Jose13}). 

For ATI, the utilization of the 3D-TDSE [Eq.~(\ref{tdse3d})] allow us to calculate not only energy-resolved photoelectron spectra, but also
  angular electron momentum distributions of atoms driven by spatially inhomogeneous fields. As in the 1D case the nonhomogeneous character of the laser electric field plays an important role on the ATI phenomenon. In addition, our 3D approach is able to model in a reliable way the ATI process both in the tunneling and multiphoton regimes. We show that for the former, the spatial nonhomogeneous field causes significant modifications on the electron momentum distributions and photoelectron spectra, while its effects in the later appear to be negligible. Indeed, through the tunneling ATI process, one can obtain higher energy electrons as well as a high degree of
asymmetry in the momentum space map. In our study we consider NIR laser fields with intensities in the mid- $10^{14}$ W cm$^{-2}$ range. We use a linear approximation for the plasmonic field, considered valid when the electron excursion is small compared with the inhomogeneity region. Indeed, our 3D simulations confirm that plasmonic fields could drive electrons with energies in the near-keV regime (see e.g.~\cite{Marcelo13A}). 


Similarly to the 1D case, the ATI spectrum is calculated starting from the time propagated electron wave function, once the laser pulse has ceased. For computing the energy-resolved photoelectron spectra $P(E)$ and two-dimensional electron distributions $\mathcal{H}(k_z,k_r)$, where $k_z$ ($k_r$) is the electron momentum component parallel (perpendicular) to the polarization direction, we use the window function approach developed in~\cite{schaferwop1,schaferwop}.

Experimentally speaking, both the direct and rescattered electrons contribute to the energy-resolved photoelectron spectra. It means that for tackling this problem both physical mechanisms should to be included in any theoretical model. In that sense, the 3D-TDSE, which can be considered as an exact approach to the ATI problem for atoms and molecules in the SAE approximation, appears to be the most suitable tool to predict the $P(E)$ in the whole range of electron energies.

\subsection{Semiclassical approach}

An independent approach to compute high-order harmonic spectra for atoms in intense laser pulses is the Strong Field Approximation (SFA) or Lewenstein model~\cite{Lewenstein94}.
The main ingredient of this approach is the evaluation of the time-dependent dipole moment $\mathbf{d}(t)$. Within the SAE approximation, it can be calculated starting from the ionization and recombination transition matrices combined with the classical action of the laser-ionized electron moving in the laser field. The SFA approximation has a direct interpretation in terms of the so-called three-step or simple man's model~\cite{Lewenstein94,corkum93}. 

Implicitly the Lewenstein model deals with spatially homogeneous electric and vector potential fields, i.e.~fields that do not experience variations in the region where the electron dynamics takes place. In order to consider spatial nonhomogeneous fields, the SFA approach needs to be modified accordingly, i.e.~the ionization and recombination transition matrices, joint with the classical action, now should take into account this new feature of the laser electric and vector potential fields (for details see~\cite{Marcelo12A,Marcelo13AA}).

As for the case of HHG driven by spatially inhomogeneous fields, ATI can also be modeled by using the SFA. In order to do so, it is necessary to modify the SFA ingredients, namely the classical action and the saddle point equations. The latter are more complex, but appear to be solvable for the case of spatially linear inhomogeneous fields (for details see~\cite{Marcelo13AA}). Within SFA it is possible to investigate how the individual pairs of quantum orbits contribute to the photoelectron spectra and the two-dimensional electron momentum distributions. We demonstrate that the quantum orbits have a very different behavior in the spatially inhomogeneous field when compared to the homogeneous field. In the case of inhomogeneous fields, the ionization and rescattering times differ between neighboring cycles, despite the field being nearly monochromatic. Indeed, the contributions from one cycle may lead to a lower cutoff, while another may develop a higher cutoff. As was shown both by our quantum mechanical and classical models, our SFA model confirms that the ATI cutoff extends far beyond the semiclassical cutoff, as a function of inhomogeneity strength. In addition, the angular momentum distributions have very different features compared to the homogeneous case. For the neighboring cycles, the electron momentum distributions do not share the same absolute momentum, and as a consequence they do not have the same yield. 

\subsection{Classical framework}
\label{classical}

Important information such as the HHG cutoff and the properties of the electron trajectories moving in the oscillatory laser electric field, can be obtained solving the classical one-dimensional Newton-Lorentz equation for an electron moving in a linearly polarized electric field. Specifically, we find the numerical solution of 
\begin{equation}
\label{newton}
\ddot{x}(t)=-\nabla_x V_{\rm{l}}(x,t),
\end{equation}
where $V_{\rm{l}}(x,t)$ is defined in Eq.~(\ref{vlaser1d}) with the laser electric field linearly polarized in the $x$ axis. For fixed values of ionization times $t_i$, it is possible to obtain the classical trajectories and to numerically calculate the times $t_r$ for which the electron recollides with the parent ion. In addition, once the ionization time $t_i$ is fixed, the full electron trajectory is completely determined (for more details about the classical model see~\cite{Marcelo15CPC}).

The following conditions are commonly set (the resulting model is also known as the simple man's model): i) the electron starts with zero velocity at the origin at time $t=t_i$, i.e., $x(t_i)=0$ and $\dot{x}(t_i)=0$; (ii) when the laser electric field reverses its direction, the electron returns to its initial position, i.e., recombines with the parent ion, at a later time, $t=t_r$, i.e. $x(t_r)=0$. $t_i$ and $t_r$ are known as ionization and recombination times, respectively. The electron kinetic energy at $t_r$ can be obtained from the usual formula $E_k(t_r)=\dot{x}(t_r)^2/2$, and, finding the value of $t_r$ (as a function of $t_i$) that maximizes this energy, we find that the HHG cutoff is given by $n_c\omega_0=3.17 U_p+I_p$, where $n_c$ is the harmonic order at the cutoff, $\omega_0$ is the laser frequency, $U_p$ is the ponderomotive energy and $I_p$ is the ionization potential of the atom or molecule under consideration. It is worth mentioning that the HHG cutoff will be extended when spatially inhomogeneous fields are employed.

\begin{figure*} [htb]
\includegraphics[width=.6\columnwidth]{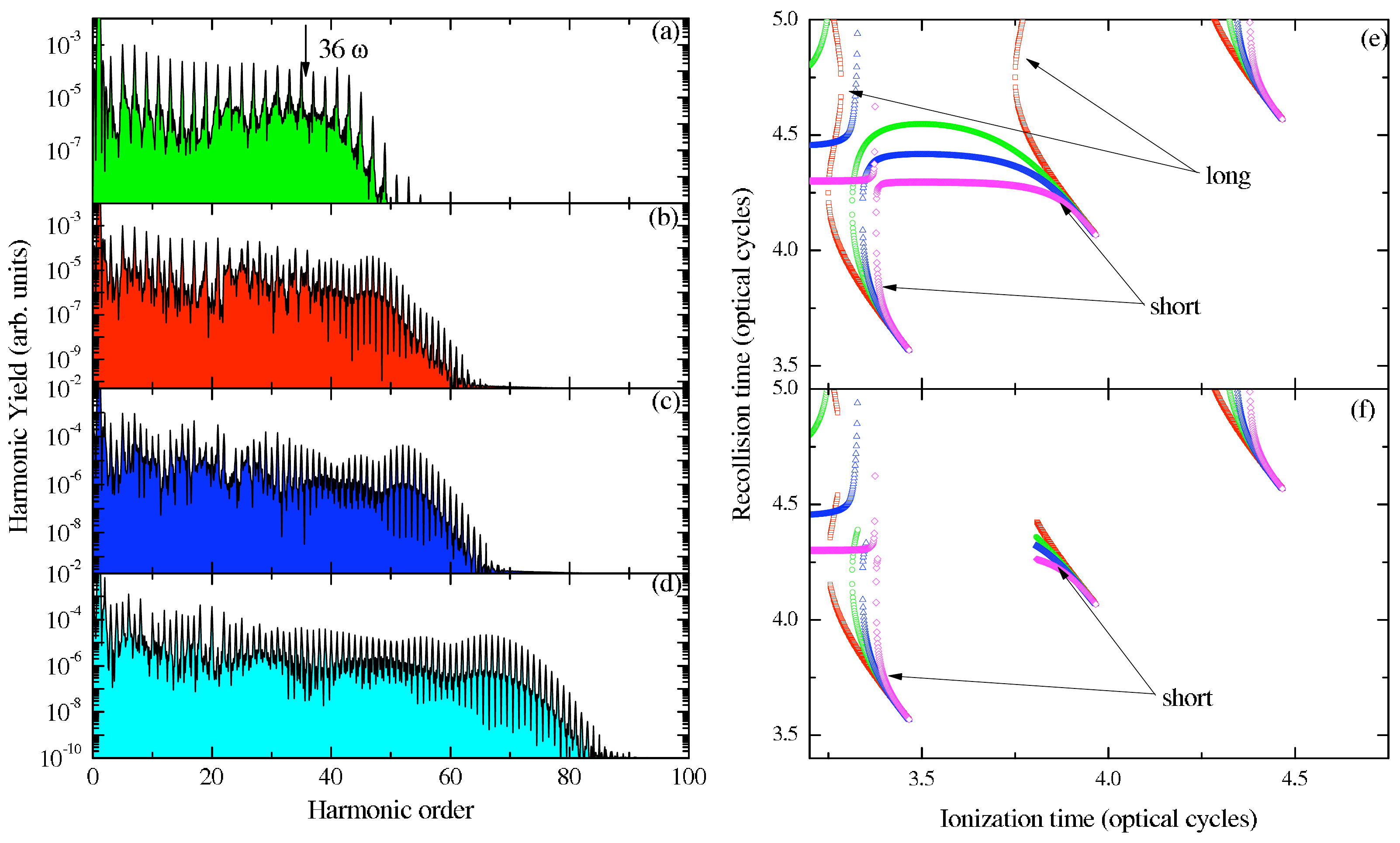}
\caption{HHG spectra for a model atom with a ground-state energy, $I_p=-0.67$ a.u. obtained using the 1D-TDSE approach. The laser parameters are $I=2\times10^{14}$ W$\cdot$cm$^{-2}$ and $\lambda=800$ nm. We have used a trapezoidal shaped pulse with two optical cycles turn on and turn off, and a plateau with six optical cycles, 10 optical cycles in total, i.e.~approximately 27 fs. The arrow indicates the cutoff predicted by the semiclassical model~\cite{Lewenstein94}. Panel (a): homogeneous case, (b): $\varepsilon=0.01$ (100 a.u), (c): $\varepsilon=0.02$ (50 a.u) and (d): $\varepsilon=0.05$ (20 a.u). The numbers in brackets indicate an estimate of the inhomogeneity region (for more details see e.g~\cite{Husakou11A,Marcelo12A}) . In panels (e) and (f) is shown the dependence of the semiclassical trajectories on the ionization and recollision times for different values of $\varepsilon$ and for the non confined case, panel (e) and the confined case, panel (f), respectively. Red squares: homogeneous case $\varepsilon=0$; green circles: $\varepsilon=0.01$; blue triangles: $\varepsilon=0.02$ and blue triangles: $\varepsilon=0.05$.}
\label{Figure1HHG}
\end{figure*}

From the simple-man's model~\cite{corkum93,Lewenstein94} we can describe the physical origin of the
ATI process as follows: an atomic electron at a position $x=0$, is released
or \textit{born} at a given ionization time $t_{i}$, 
with zero velocity, i.e.~$\dot{x}(t_{i})=0$. This electron now moves
only under the influence of the oscillating laser electric field (the
residual Coulomb interaction is neglected in this model) and will reach
the detector either directly or through a rescattering process.  By using the
classical equation of motion, it is possible to calculate the maximum
energy of the electron for both direct and rescattered processes.

For the direct ionization, the kinetic energy of an electron released or born at time $t_{i}$ is
\begin{equation}
\label{direct}
E_{d}=\frac{\left[ \dot{x}(t_{i})-\dot{x}(t_{f})\right] ^{2}}{2},
\end{equation}
where $t_{f}$ is the end time of the laser pulse. For the rescattering process, in which the electron returns to the core at a time $t_{r}$ and reverses its direction, the kinetic energy  of the electron yields
\begin{equation}
\label{rescattered}
E_{r}=\frac{\left[ \dot{x}(t_{i})+\dot{x}(t_{f})-2\dot{x}(t_{r})\right] ^{2}}{2}.
\end{equation}

For homogeneous fields, Eqs.~(\ref{direct}) and (\ref{rescattered}) become $E_{d}=\frac{\left[ A(t_{i})-A(t_{f})\right] ^{2}}{2}$ and 
$E_{r}=\frac{\left[ A(t_{i})+A(t_{f})-2A(t_{r})\right] ^{2}}{2}$, with $A(t)$ being the laser vector potential $A(t)=-\int^{t} E(t')dt'$. For the case with $\varepsilon=0$, it can be shown that the maximum value for $E_{d}$ is $2U_{p}$ while for $E_{r}$ it is $10U_{p}$~\cite{Milosevic06}. These two values appear as cutoffs in the energy-resolved photoelectron spectrum. 


%

\section{HHG driven by spatially inhomogeneous fields}

Field-enhanced high-order-harmonic generation (HHG) using plasmonics fields, generated starting from engineered nanostructures or nanoparticles, requires no extra amplification stages due to the fact that, by exploiting surface plasmon resonances, the input driving electric field can be enhanced by more than 20 dB (corresponding to an increase in the intensity of 2-3 orders of magnitude). As a consequence of this enhancement, the threshold laser intensity for HHG generation in noble gases is largely exceeded and the pulse repetition rate remains unaltered. In addition, the high-harmonics radiation generated from each nanosystem acts as a pointlike source, enabling a high collimation or focusing of this coherent radiation by means of (constructive) interference. This fact opens a wide range of possibilities to spatially arrange nanostructures to enhance or shape the spectral and spatial properties of the harmonic radiation in numerous ways~\cite{Kim08,Park11,Kovacev13NJP}.

Due to the nanometric size of the so-called plasmonic 'hot spots', i.e.~the spatial region where the electric field reaches its highest intensity, one of the main theoretical assumptions, namely the spatial homogeneity of the driven electric field, should be excluded. As a consequence, both the analytical and numerical approaches to study laser-matter processes in atoms and molecules, in particular HHG, need to be modified to treat adequately this different scenario and allow now for a spatial dependence in the laser electric field. Several authors have addressed this problem recently~\cite{AlexisVG,Yavuz15,Husakou11A,Husakou11OE,Marcelo12OE,Marcelo12A,Marcelo12AA,Marcelo12AAA,Marcelo12JMO,Yavuz12,Jose13, Yavuz13,Marcelo13A,Marcelo13AR,Marcelo13AP,Marcelo13AA,Marcelo13LPL,Marcelo14,Marcelo14EPJD, Marcelo15CPC,Marcelo15,Husakou14A,Ebadi14A,Fetic12,Luo13,Feng13,Wang13,Lu13A,He13,Zhang13,Luo13JOSAB, Cao14,Cao14a,Feng15,Yu15}. As we will show below, this new characteristic affects considerably the electron dynamics and this is reflected on the observables, in the case of this subsection the HHG spectra.

\subsection{Spatially (linear) nonhomogeneous fields and electron confinement}

\begin{figure} [htb]
\includegraphics[width=\columnwidth]{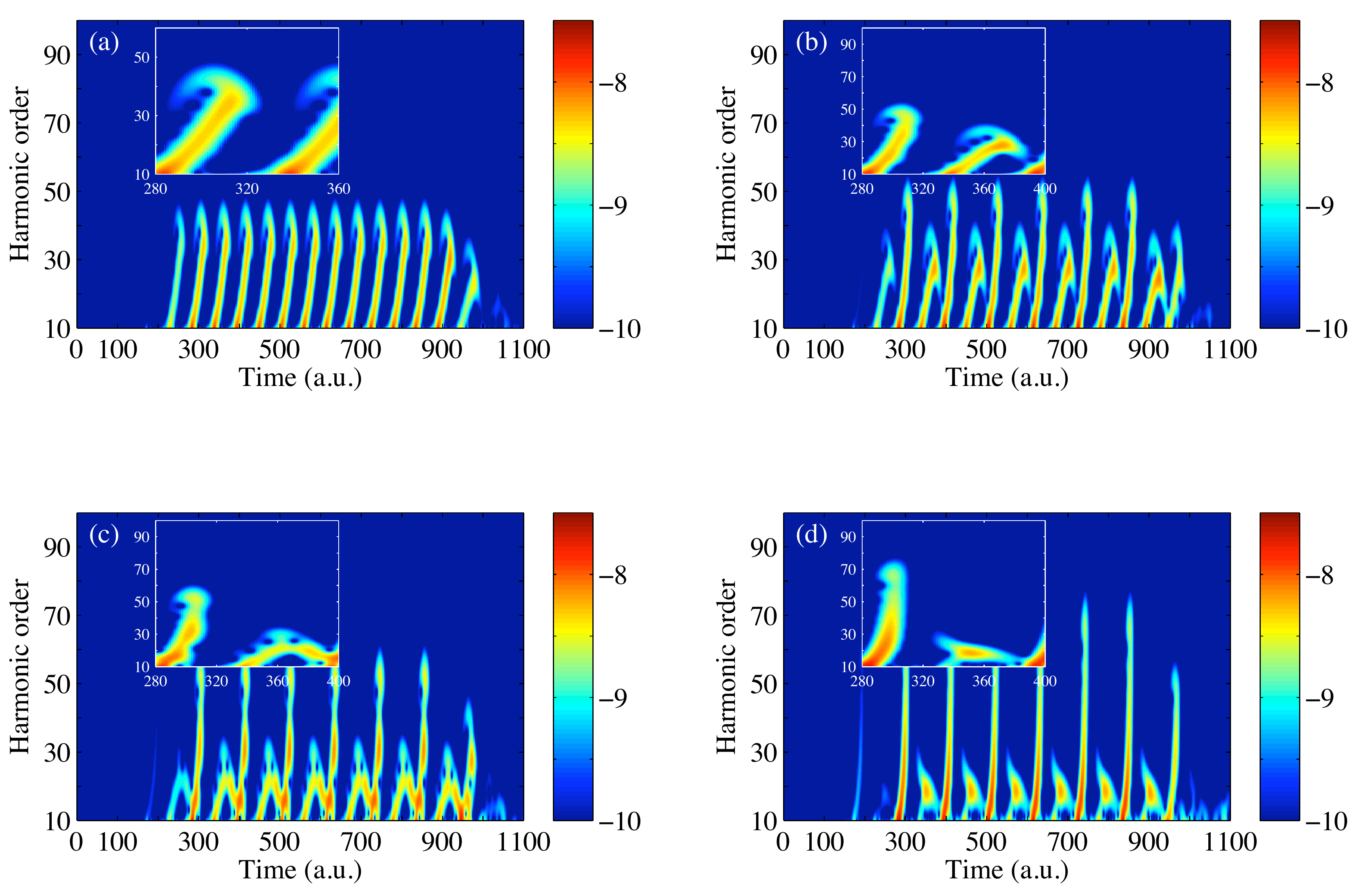}
\caption{Panels (a)-(d): Gabor analysis for the corresponding HHG spectra of panels (a)-(d) of Fig.~\ref{Figure1HHG}. The zoomed regions in all panels show a time interval during the laser pulse for which the complete electron trajectory, from birth time to recollision time, falls within the pulse plateau. In panels (a)-(d) the color scale is logarithmic.}
\label{Figure2HHG}
\end{figure}

In this sub-section we summarize the study carried out in~\cite{Marcelo12A} where it is demonstrated that both the inhomogeneity of the local fields and the constraints in the electron movement, play an important role in the HHG process and lead to the generation of even harmonics and a significant increase in the HHG cutoff, more pronounced for longer wavelengths. In order to understand and characterize these new HHG features we employ two of the different approaches mentioned above: the numerical solution of the 1D-TDSE (see panels (a)-(d) in Fig.~\ref{Figure1HHG}) and the semiclassical approach known as Strong Field Approximation (SFA). Both approaches predict comparable results and describe satisfactorily the new features, but by employing the semiclassical arguments (see panels (e), (f) in Fig.~\ref{Figure1HHG}) behind the SFA and time-frequency analysis tools (Fig.~\ref{Figure2HHG}), we are able to fully explain the reasons of the cutoff extension.

\newpage

\subsection{Spatially (linear) nonhomogeneous fields: the SFA approach}

In this subsection we summarize the work presented in \cite{Marcelo12AA}. In this contribution, we perform a detailed analysis of high-order harmonic generation (HHG) in atoms within the strong field approximation (SFA) by considering spatially (linear) inhomogeneous monochromatic laser fields. We investigate how the
individual pairs of quantum orbits contribute to the harmonic spectra. To this end we have modified both the classical action and the saddle points equations by including explicitly the spatial dependence of the laser field. We show that in the case of a linear inhomogeneous field the electron tunnels with two different canonical momenta. One of these momenta leads to a higher cutoff and the other one develops a lower cutoff. Furthermore, we demonstrate that the quantum orbits have
a very different behavior in comparison to the conventional homogeneous field. A recent study supports our initial findings~\cite{carla2016}.

We also conclude that in the case of the inhomogeneous fields both odd and even harmonics are present in the HHG spectra. Within our extended SFA model, we show that the HHG cutoff extends far beyond the standard semiclassical cutoff in spatially homogeneous fields. Our findings are in good agreement both with quantum-mechanical and classical models. Furthermore, our approach confirms the versatility of the SFA approach to tackle now the HHG driven by spatially (linear) inhomogeneous fields. 

\begin{figure}[htb]
\centering
\includegraphics[width=\columnwidth]{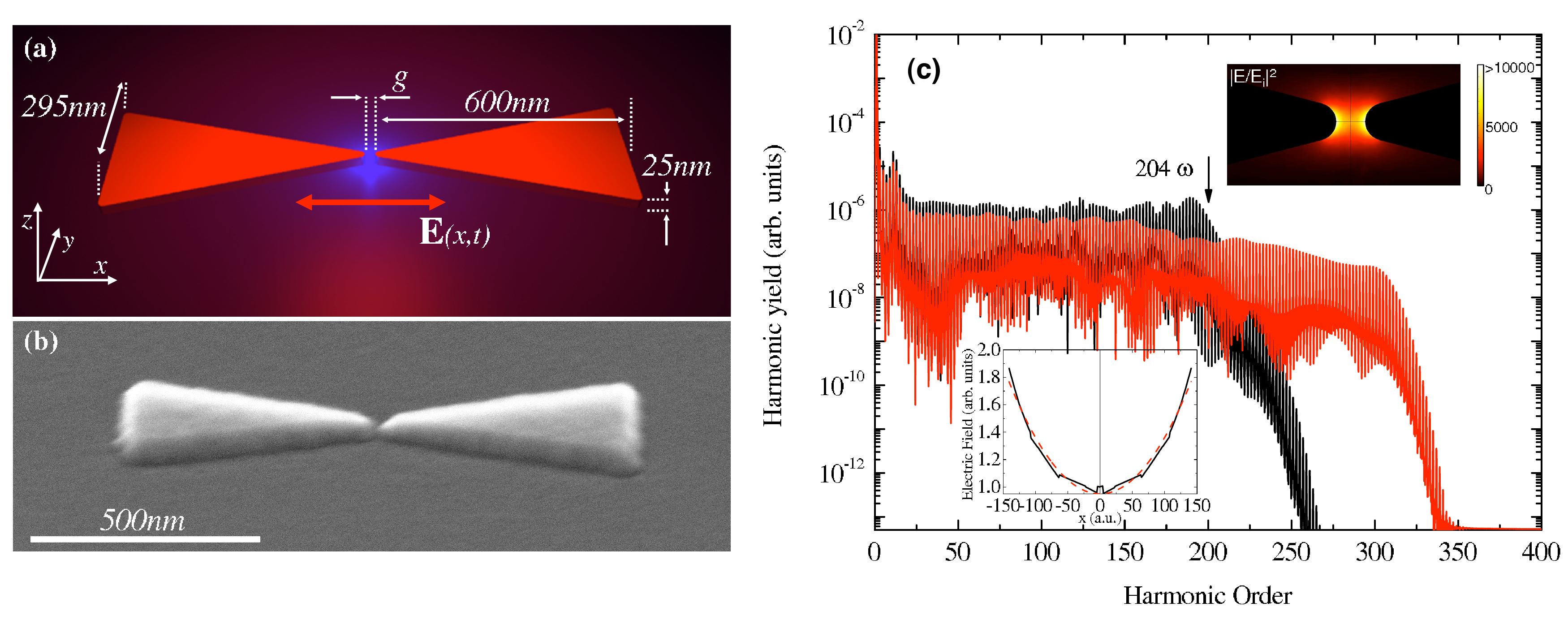}
\caption{(a) Schematic representation of the geometry of the
considered nanostructure. A gold bow-tie antenna resides on glass substrate
(refractive index $n = 1.52$) with superstate medium of air ($n = 1$). The
characteristic dimensions of the system and the coordinate system used in
the 1D-TDSE simulations are shown. (b) SEM image of a real gold bow-tie antenna. (c) High-order harmonic generation (HHG) spectra for a model of argon atoms ($I_p=-0.58$ a.u.), driven by a laser pulse with wavelength $\protect\lambda=1800$ nm and intensity $I=1.25\times10^{14}$ W$\cdot$cm$^{-2}$ at the center of the gap $x=0$. We have used a trapezoidal shaped pulse with three optical cycles turn on and turn off, and a plateau with four optical cycles (about 60 fs). The gold bow-tie nanostructure has a gap $g=15$ nm (283 a.u.). The black line indicates the homogeneous case while the red line indicates the nonhomogeneous case. The arrow indicates the cutoff predicted by the semiclassical model
for the homogeneous case~\cite{Lewenstein94}. The top left inset shows the
functional form of the electric field $E(x,t)$, where the solid lines are the
raw data obtained from the finite element simulations and the dashed line is a
nonlinear fitting. The top right inset shows the intensity enhancement in
the gap region of the gold bow-tie nanostructure. }
\label{Figure3HHG}
\end{figure}

\subsection{Real nonhomogeneous fields}

In this sub-section we present numerical simulations of HHG in an argon model atom produced by the fields generated when a gold bow-tie nanostructure is illuminated by a short laser pulse of long wavelength $\lambda=1800$ nm (see~\cite{Marcelo12OE} for more details). The functional form of these fields is extracted from finite element simulations using both the complete geometry of the metal nanostructure and laser pulse characteristics (see Fig.~\ref{Figure3HHG}(a)).  We use the numerical solution of the TDSE in reduced dimensions to predict the HHG spectra. A clear extension in the harmonic cutoff position is observed. This characteristic could lead to the production of XUV coherent laser sources and open the avenue to the generation of shorter attosecond pulses. It is shown in Fig.~\ref{Figure3HHG}(c) that this new feature is a consequence of the combination of a spatial nonhomogeneous electric field, which modifies substantially the electron trajectories, and the confinement of the electron dynamics. Furthermore, our numerical results are supported by time-analysis
and classical simulations. A more pronounced increase in the harmonic cutoff, in addition to an appreciable growth in conversion efficiency,
could be attained by optimizing the nanostructure geometry and materials. These degrees of freedom could pave the way to tailor the
harmonic spectra according to specific requirements.

\subsection{Temporal and spatial synthesized fields}

In this sub-section we present a brief summary of the results published in~\cite{Jose13}. In short, numerical simulations of HHG in He atoms using a temporal and spatial synthesized laser field are considered using the full 3D-TDSE. This particular field provides a new route for the generation of photons at energies beyond the carbon K-edge using laser pulses at 800 nm, which can be obtained from conventional Ti:Sapphire laser sources.  The temporal synthesis is performed using two few-cycle laser pulses delayed in time~\cite{Jose09A}. On the other hand, the spatial synthesis is obtained by using a spatial nonhomogeneous laser field~\cite{Husakou11A,Yavuz12, Marcelo12A} produced when a laser beam is focused in the vicinity of a metal nanostructure or nanoparticle.

Focusing on the spatial synthesis, the nonhomogeneous spatial distribution of the laser electric field can be obtained experimentally by using the resulting field as produced after the interaction of the laser pulse with nanoplasmonic antennas~\cite{Husakou11A, Yavuz12, Marcelo12A,Kim08}, metallic nanowaveguides~\cite{Park11}, metal~\cite{Zherebtsov11,SuessmannSPIE11} and dielectric nanoparticles~\cite{SuessmannPRB11} or metal nanotips~\cite{PeterH06,Schenk10,Kruger11,Kruger12N,Kruger12B,Herink12}. 

The coupling between the atom and the laser pulse, linearly polarized along the $z$ axis, is modified in order to treat the spatially nonhomogeneous fields and can be written it as: $V_{\rm{l}}(z,t,\tau)=\tilde{E}(z,t,\tau)\,z$ with $\tilde{E}(z,t,\tau)=E(t,\tau)(1+\varepsilon z)$ and $E(t,\tau)=E_1(t)+E_2(t,\tau)$ the temporal synthesized laser field with $\tau$ the time delay between the two pulses (see e.g.~\cite{Jose09A} for more details). As in the 1D case the parameter $\varepsilon$ defines the strength of the nonhomogeneity and the dipole approximation is preserved because $\varepsilon\ll1$.

\begin{figure}[htb]
\centering
\includegraphics[width=\columnwidth]{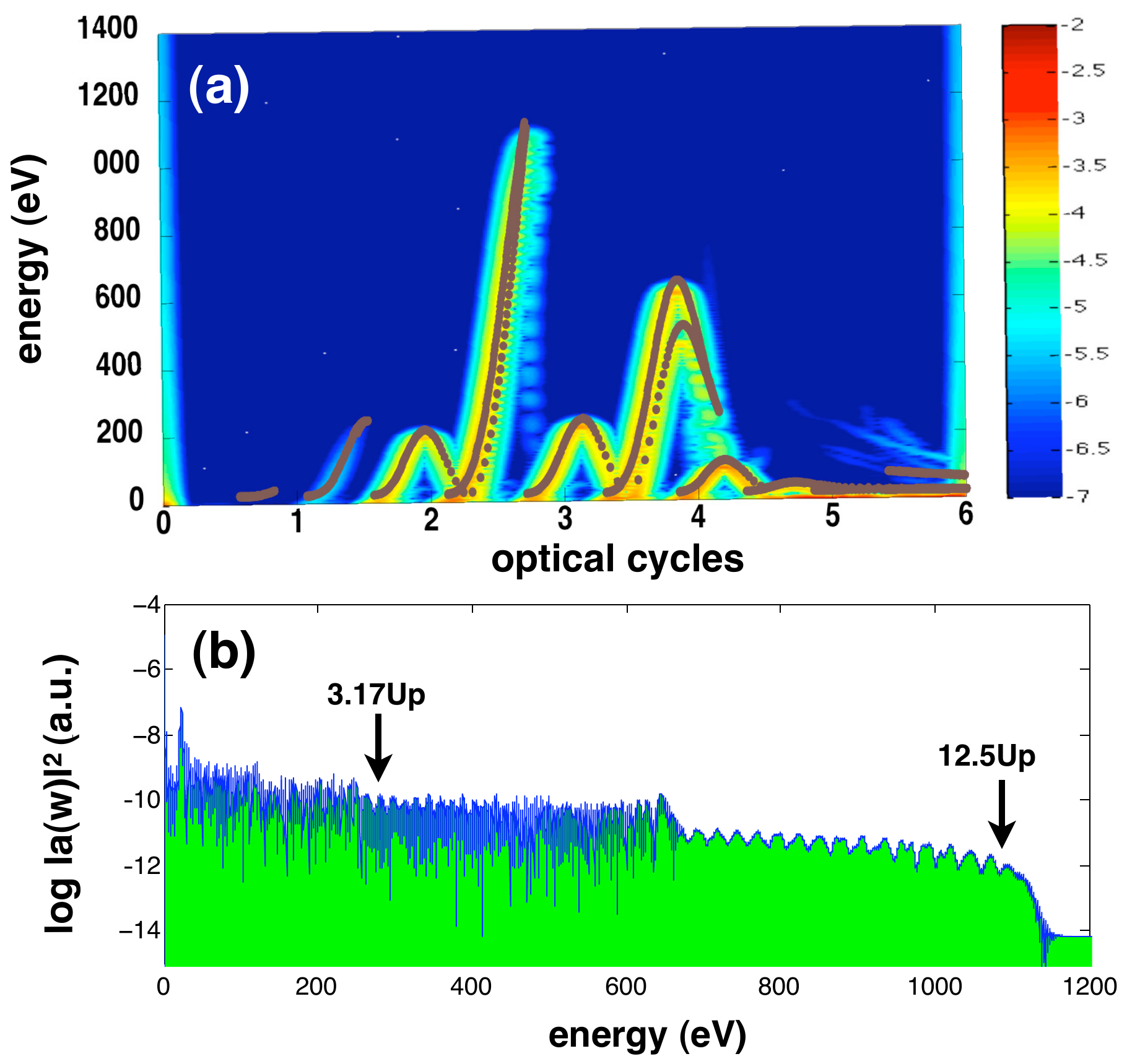}
\caption{(a) Time-frequency analysis obtained from the 3D-TDSE harmonic spectrum for a He atom driven by the spatially and temporally synthesized pulse described in the text with $\varepsilon=0.002$. The plasmonic enhanced intensity $I=1.4\times 10^{15}$ W cm$^{-2}$. Superimposed (in brown) are the classical rescattering energies; (b) 3D-TDSE harmonic spectrum for the same parameters used in (a).}.
\label{Figure4HHG}
\end{figure}

The linear functional form for the spatial non-homogeneity described above could be obtained engineering adequately the geometry of plasmonic nanostructures and by adjusting the laser parameters in such a way that the laser-ionized electron feels only a linear spatial variation of the laser electric field when in the continuum  (see e.g.~\cite{Marcelo12OE} and references therein). The harmonic spectrum then obtained in He for $\varepsilon=0.002$ is presented in Fig.~\ref{Figure4HHG}(b).  We can observe a considerable cut-off extension up to $12.5 U_p$ which is much larger when compared with the double pulse configuration employed alone (it leads only to a maximum of $4.5 U_p$~\cite{Jose09A}). This large extension of the cutoff is therefore a signature of the combined effect of the double pulse and the spatial nonhomogeneous character of the laser electric field. For this particular value of the laser peak intensity ($1.4\times 10^{15}$ W cm$^{-2}$) the highest photon energy is greater than 1 keV. Note that the quoted intensity is actually the plasmonic enhanced intensity, not the input laser intensity. The latter could be several orders of magnitude smaller, according to the plasmonic enhancement factor (see e.g.~\cite{Kim08,Park11}) and will allow the nanoplasmonic target to survive to the interaction.  In order to confirm the underlying physics highlighted by the classical trajectories analysis, we have retrieved the time-frequency distribution of the calculated dipole (from the 3D-TDSE) corresponding to the case of the spectra presented in Fig.~\ref{Figure4HHG}(b) using a wavelet analysis. The result is presented in Fig.~\ref{Figure4HHG}(a) where we have superimposed the calculated classical recombination energies (in brown) to show the excellent agreement between the two theoretical approaches. The consistency of the classical calculations with the full quantum approach is clear and confirms the mechanism of the generation of this $12.5 U_p$ cut-off extension. In addition, the HHG spectra exhibit a clean continuum as a result of the trajectory selection on the recombination time, which itself is a consequence of employing a combination of temporally and spatially synthesized laser field.

\subsection{Plasmonic near-fields}
\label{HHGnearfield}

This sub-section includes an overview of the results reported in~\cite{Marcelo13AR}. In this contribution it is shown how the HHG spectra from model Xe atoms are modified by using a plasmonic near enhanced field generated when a metal nanoparticle is illuminated by a short laser pulse. A setup combining a noble gas as a driven media and metal nanoparticles was also proposed recently in~\cite{Husakou14A,Husakou15}. 

For our near-field we use the function given by~\cite{SuessmannSPIE11} to define the spatial nonhomogeneous laser electric field $E(x,t)$, i.e.
\begin{equation}
E(x,t)=E_{0}\,f(t)\,\exp (-x/\chi )\sin (\omega_0 t+\phi),
\end{equation}
where $E_{0}$, $\omega_0$, $f(t)$ and $\phi $ are the peak amplitude, the laser field frequency, the field envelope and the CEP, respectively. The functional form of the resulting laser electric field is extracted from attosecond streaking experiments and incorporated both in our quantum and classical approaches. In this specific case the spatial dependence of the plasmonic near-field is given by $\exp(-x/\chi )$ and it is a function of both the size and the material of the spherical nanoparticle. $E(x,t)$ is valid for $x$ outside of the metal nanoparticle, i.e.~$x\ge R_0$, where $R_0$ is its radius.  It is important to note that the electron motion takes place in the region $x\ge R_0$ with $(x+R_0)\gg0$. We consider the laser field having a sin$^{2}$ envelope: $f(t)=\sin ^{2}\left( \frac{\omega_0 t}{2n_{p}}\right)$, where $n_{p}$ is the total number of optical cycles, i.e.~the total pulse duration is $\tau_L =2\pi n_{p}/\omega_0$. The harmonic yield of the atom is obtained by Fourier transforming the acceleration $a(t)$ of the electronic wavepacket.

\begin{figure}[htb]
\centering
\includegraphics[width=\columnwidth]{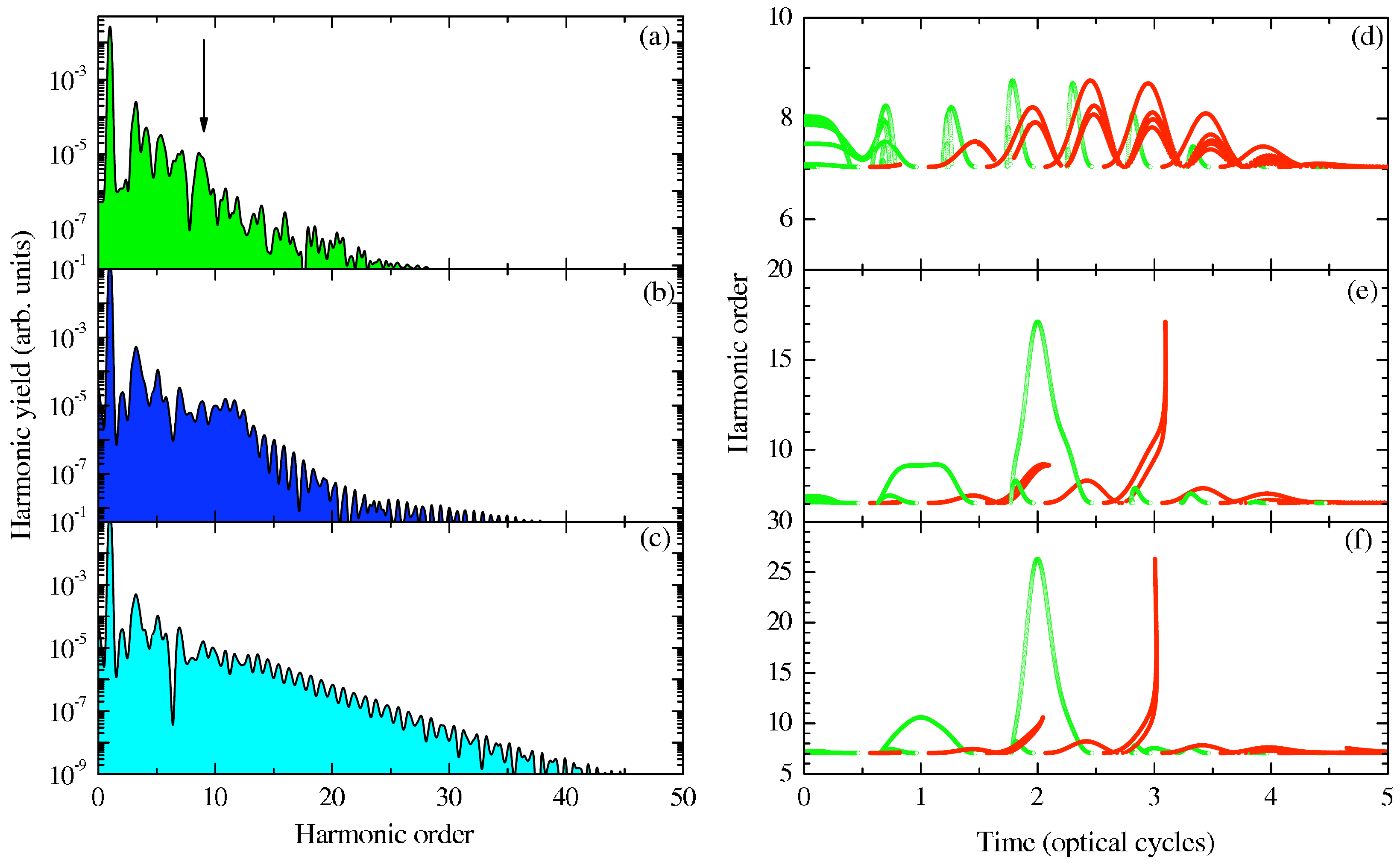}
\caption{HHG spectra for model Xe atoms, laser wavelength $\protect\lambda=720$ nm
and intensity $I=2\times10^{13}$ W$\cdot$cm$^{-2}$. We use a sin$^{2}$ pulse envelope with $n=5$. Panel (a) represents the homogeneous case, panel (b) $\protect\chi=50$ and panel (c) $\protect\chi=40$. The arrow in panel (a) indicates the cutoff predicted by the semiclassical approach~\cite{Lewenstein94}. Panels (d), (e), (f) show the corresponding total energy of the electron (expressed in harmonic order) driven by the laser field calculated from the one-dimensional Newton-Lorentz equation and plotted as a function of the $t_i$ (green (light gray) circles) or the $t_r$ (red (dark gray) circles).}
\label{Figure5HHG}
\end{figure}

Figure~\ref{Figure5HHG},  panels (a), (b) and (c) show the harmonic spectra for model xenon atoms generated by a laser pulse with $I=2\times 10^{13}$ W cm$^{-2}$, $\lambda =720$ nm and a $\tau
_L=13$ fs, i.e.~$n_{p}=5$ (which corresponds to an intensity envelope of $\approx4.7$ fs FWHM)~\cite{SuessmannSPIE11}. 
In the case of a spatial homogeneous field, no harmonics beyond the $9^{th}$ order are observed. The spatial decay parameter $\chi$ accounts for the spatial nonhomogeneity induced by the nanoparticle and it varies together with its size and the kind of metal employed. Varying  the value of $\chi$ is therefore equivalent to choosing the type of nanoparticle used, which allows to overcome the semiclassically predicted cutoff limit and reach higher harmonic orders. For example, with $\chi =40$ and $\chi=50$ harmonics in the mid 20s (panel c) and well above the $9^{th}$ (a clear cutoff at $n_c\approx 15$ is achieved) (panel b), respectively, are obtained. A modification in the harmonic periodicity, related to the breaking of symmetry imposed by the induced nonhomogeneity, is also clearly noticeable.

Now, by the semiclassical simple man's (SM) model~\cite{corkum93,Lewenstein94} we will study the harmonic cut-off extension. This new effect may be caused by a combination of several factors (for details see~\cite{Marcelo12A,Marcelo12OE}). As is well known, the cutoff law is $n_{c}=(3.17U_{p}+I_{p})/\omega_0$, where $n_{c}$ is the harmonic order at the cutoff and $U_{p}$ the ponderomotive energy. We solve numerically Eq.~(\ref{newton}) for an electron moving in an electric field with the same parameters used in the TDSE-1D calculations, i.e.
\begin{equation}
\ddot{x}(t) =-\nabla _{x}V_{l}(x,t) = -E(x,t)\left(1-\frac{x(t)}{\chi }\right),
\end{equation}
and consider the SM model initial conditions: the electron starts at position zero at $t=t_{i}$ (the ionization time) with zero velocity, i.e. $x(t_{i})=0$ and $\dot{x}(t_{i})=0$. When the electric field reverses, the electron returns to its initial position (i.e.~the electron \textit{recollides} or recombines with the parent ion) at a later time $t=t_{r}$ (the recombination time), i.e.~$x(t_{r})=0$. The electron kinetic energy at the $t_{r}$ is calculated as usual from: $E_{k}(t_{r})=\frac{\dot{x}(t_{r})^{2}}{2}$ and finding the $t_r$ (as a function of $t_{i}$) that maximizes $E_k$, $n_c$ is also maximized.

Panels (d), (e) and (f) of Fig.~\ref{Figure5HHG} represent the behaviour of the harmonic order upon the $t_{i}$ and $t_{r}$, calculated from $n=(E_{k}(t_{i,r})+I_{p})/ \omega_0 $ as for the cases (a), (b) and (c) of Fig.~\ref{Figure5HHG}, respectively. Panels (e) and (f) show how the nonhomogeneous character of the laser field strongly modifies the electron trajectories towards an extension of the $n_c$. This is clearly present at $n_{c}\sim 18\omega_0$ (28 eV) and $n_{c}\sim 27\omega_0$ (42 eV) for $\chi =50$ and $\chi =40$, respectively. These last two cutoff extensions are consistent with the quantum predictions presented in panels (b) and (c) of Fig.~\ref{Figure5HHG}.

Classical and quantum approaches predict cutoff extensions that could lead to the production of XUV coherent laser sources and open a direct route to the generation of attosecond pulses.  This effect is caused by the induced laser field spatial nonhomogeneity, which modifies substantially the electron trajectories. A more pronounced increment in the harmonic cutoff, in addition to an appreciable growth in the conversion efficiency, could be reached by varying both the radius and the metal material of the spherical nanoparticles. These new degrees of freedom could pave the way to extend the harmonic plateau reaching the XUV regime with modest input laser intensities.

\section{ATI driven by spatially inhomogeneous fields}

As was mentioned at the outset, ATI represents another fundamental strong field phenomenon.  Investigations carried out on ATI, generated by few-cycle driving laser pulses, have attracted much interest due to the sensitivity of the energy and angle-resolved photoelectron spectra to the absolute value of the CEP~\cite{Milosevic06,paulus2011}. This feature makes the ATI phenomenon a conceivable tool for laser pulse characterization. In order to characterize the CEP of a few-cycle laser pulse, the so-called backward-forward asymmetry of the ATI spectrum is measured and from the information collected the absolute CEP value can be obtained~\cite{paulus2001,paulus2011}. Furthermore, nothing but the high energy region of the photoelectron spectrum appears to be strongly sensitive to the absolute CEP and consequently electrons with high kinetic energy are needed in order to describe it~\cite{Milosevic06,paulus2001,paulus_measurement_2003}.

Nowadays, experiments have demonstrated that ATI photoelectron spectra could be extended further by using plasmon-field enhancement~\cite{Kim08,Zherebtsov11}. The strong confinement of the plasmonics spots and the distortion of the electric field by the surface plasmons induces a spatial inhomogeneity in the driving laser field, just before the interaction with the corresponding target gas. A related process employing solid state targets instead of atoms and molecules in gas phase is the so called above-threshold photoemission (ATP). This laser driven phenomenon has received special attention recently due to its novelty and the new physics involved. In ATP electrons are emitted directly from metallic surfaces or metal nanotips and they present distinct characteristics, namely higher energies, far beyond the usual cutoff for noble gases and consequently the possibility to reach similar electron energies with smaller laser intensities (see e.g.~\cite{PeterH06,Schenk10,Kruger11,Kruger14,Herink12}). Furthermore, the photoelectrons emitted from these nanosources are sensitive to the CEP and consequently it plays an important role in the angle and energy resolved photoelectron spectra~\cite{apolonski,Zherebtsov11,Kruger11}. 



\subsection{ATI driven by spatially linear inhomogeneous fields: the 1D-case}

\begin{figure}[htb]
\includegraphics[width=\textwidth]{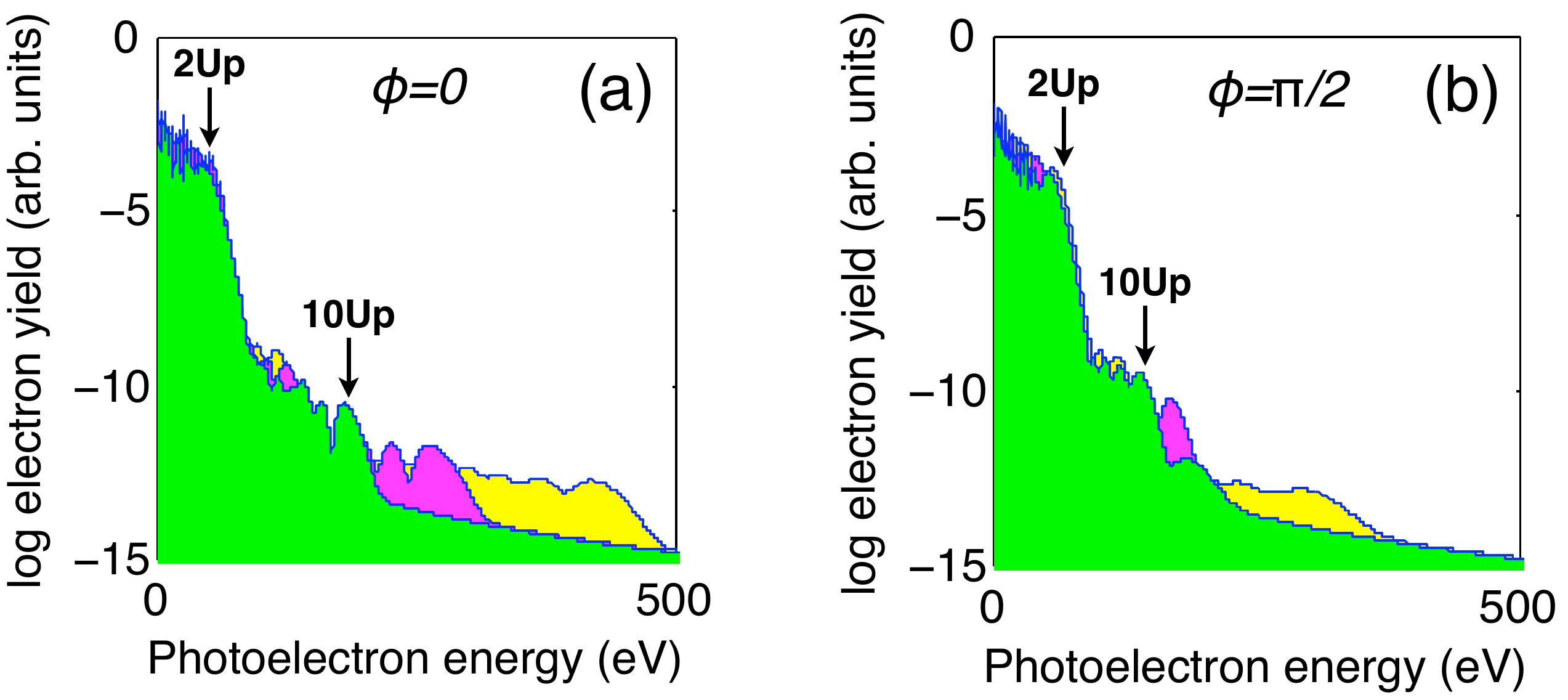}
\caption{1D-TDSE energy-resolved photoelectron spectra for a model atom with $I_{p}=-0.5$ a.u. and for the laser parameters, $I=3\times10^{14}$ W cm$^{-2}$, $\lambda=800$ nm and a sin-squared shaped pulse with a total duration of 4 cycles (10 fs). In green for $\varepsilon=0$ (homogeneous case), in magenta for $\varepsilon=0.003$ and in yellow for $\varepsilon=0.005$. Panel (a) represent the case for $\phi=0$ (sin-like pulse) and panel (b) represents the case for $\phi=\pi/2$ (cos-like pulse). The arrows indicate the $2 U_p$ and $10 U_p$ cutoffs predicted by the classical model~\cite{Milosevic06}}
\label{Figure1ATI}
\end{figure}

For our 1D quantum simulations we employ as a driving field a four-cycle (total duration 10 fs) sin-squared laser pulse with an intensity  $I=3\times10^{14}$ W cm$^{-2}$ and wavelength $\lambda=800$ nm. We chose a linear inhomogeneous field and three different values for the parameter that characterizes the
inhomogeneity strength, namely $\varepsilon =0$ (homogeneous case), $\varepsilon=0.003$ and $\varepsilon=0.005$.  Figure \ref{Figure1ATI}(a)
shows the cases with $\phi =0$ (a sin-like laser pulse) meanwhile in Fig.~\ref{Figure1ATI}(b) $\phi =\pi/2$
(a cos-like laser pulse), respectively. In both panels
green represents the homogeneous case, i.e.~$\varepsilon =0$,
magenta is for $\varepsilon =0.003$ and yellow is for $\varepsilon =0.005$,
respectively.

For the homogeneous case, the spectra exhibits the usual distinct behavior,
namely the $2U_{p}$ cutoff ($\approx 36$ eV for our case) and the $10U_{p}$
cutoff ($\approx 180$ eV), where $U_{p}=E_{0}^{2}/4\omega_0^{2}$ is the
ponderomotive potential.  The former cutoff corresponds to those electrons
that, once ionized,  never return to the atomic core, while the latter one
corresponds to the electrons that, once ionized, return to the core and
elastically rescatter. It is well established using classical arguments that the maximum kinetic energies of the \textit{direct} and the \textit{rescattered} electrons are $E_{max}^{d}=2U_{p}$ and  $E_{max}^{r}=10U_{p}$,
respectively. In a quantum mechanical approach, however, it is possible to find electrons with energies beyond the 10$U_p$, although their yield drops several orders of magnitude~\cite{Milosevic06}. 
The TDSE, which can be considered as an exact approach to the problem, is able to predict the $P(E)$ for the whole range of electron energies. In addition, the most energetic electrons, i.e.~those with $E_{k}\gg 2U_{p}$, are used to characterize the CEP of few-cycle pulses. As a result, a correct description
of the rescattering mechanism is needed.

For the spatial inhomogeneous case, the cutoff positions of both the \textit{direct} and the \textit{rescattered} electrons are extended towards larger energies.
For the \textit{rescattered} electrons, this extension  is very prominent. In
fact, for $\varepsilon =0.003$ and  $\varepsilon =0.005,$ it reaches  $\approx 260$ eV and  $\approx 420$ eV, respectively (see Fig.~\ref{Figure1ATI}(a)). 
Furthermore, it appears that the high energy region of $P(E)$, for instance,  the region between $200-400$ eV for $\varepsilon =0.005$
(Fig.~\ref{Figure1ATI} in yellow), is strongly sensitive to the CEP. This feature
indicates that the high energy region of the photoelectron spectra could resemble a new and better CEP characterization tool. It should be, however,
complemented by other well known and established CEP characterization tools, as, for instance, the forward-backward asymmetry (see~\cite{Milosevic06}). 
Furthermore, the utilization of nonhomogeneous fields would open the avenue for the
production of high energy electrons, reaching the keV regime, if a reliable control of the spatial and temporal shape of the laser electric field is attained. 

We now concentrate our efforts on explaining the extension of the
energy-resolved photoelectron spectra using classical arguments. From the
simple-man's model~\cite{corkum93,Lewenstein94} we can describe the physical origin of the
ATI process as follows: an atomic electron at a position $x=0$, is released
or \textit{born} at a given time, that we call \textit{ionization} time $t_{i}$, 
with zero velocity, i.e.~$\dot{x}(t_{i})=0$. This electron now moves
only under the influence of the oscillating laser electric field (the
residual Coulomb interaction is neglected in this model) and will reach
the detector either directly or through a rescattering process.  By using the
classical equation of motion, it is possible to calculate the maximum
energy of the electron for both direct and rescattered processes. The Newton
equation of motion for the electron in the laser field can be written as (see Eq.~(\ref{newton})):
\begin{eqnarray}
\ddot{x}(t) &=&-\nabla _{x}V_{\rm{l}}(x,t)  \notag  \label{newtonati} \\
&=&E(x,t)+\left[ \nabla _{x}E(x,t)\right] x  \notag \\
&=&E(t)(1+2\varepsilon x(t)),
\end{eqnarray}
where we have collected the time dependent part of the electric field in $E(t)$, i.e.~$E(t)=E_{0}f(t)\sin (\omega_0 t+\phi )$ and 
particularized to the case $h(x)=x$. In the limit where $\varepsilon =0$ in Eq.~(\ref{newtonati}), we recover the spatial homogeneous case. Using the classical formalism described in Section~\ref{classical}. we find the maximum energy for both the direct and rescattered electrons. As can be seen, the electron energy cutoffs now exceed the ones obtained for conventional fields (see panels (a) and (b), in green, in Fig.~\ref{Figure1ATI} and the respective arrows).

%

\begin{figure}[htb]
\centering
\includegraphics[width=\textwidth]{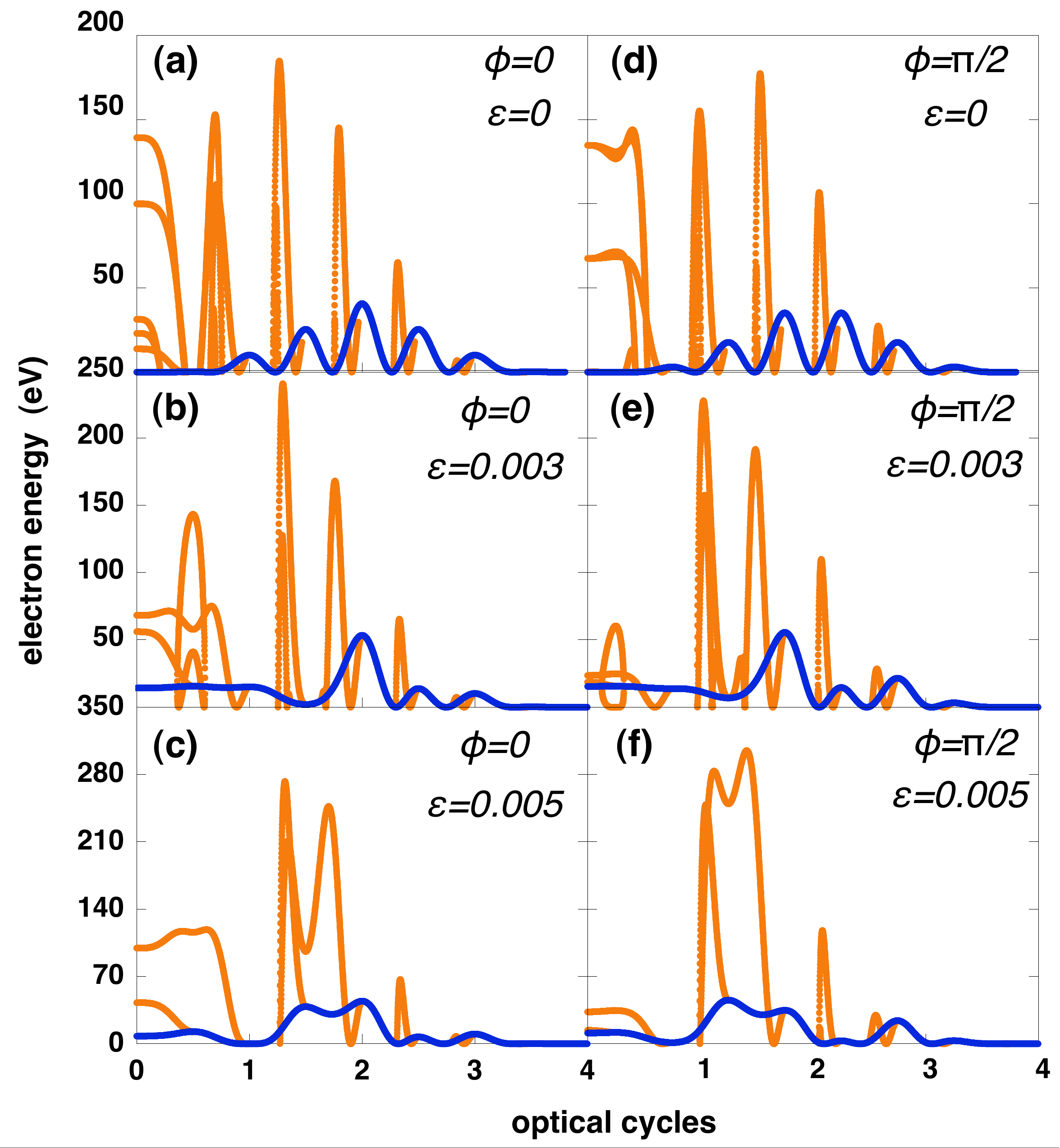}
\caption{Numerical solutions of Eq.~(\ref{newtonati}) plotted in terms of the direct (blue) and rescattered (orange) electron kinetic energy. The laser parameters are the same as in Fig.~\ref{Figure1ATI}.  Panels (a), (b) and (c) correspond to the case of sin-like pulses ($\phi=0$) and for $\varepsilon=0$ (homogeneous case), $\varepsilon=0.003$ and $\varepsilon=0.005$, respectively. Panels (d), (e) and (f) correspond to the case of cos-like pulses ($\phi=\pi/2$) and for $\varepsilon=0$ (homogeneous case), $\varepsilon=0.003$ and $\varepsilon=0.005$, respectively.}
\label{Figure2ATI}
\end{figure}

In Fig.~\ref{Figure2ATI}, we present the numerical solutions of Eq.~(\ref{newtonati}), which is plotted in terms of the kinetic energy of the direct and rescattered electrons. We employ the same laser parameters as in Fig.~\ref{Figure1ATI}. Panels (a), (b) and (c) correspond to the case of $\phi=0$ (sin-like pulses) and
for $\varepsilon=0$ (homogeneous case), $\varepsilon=0.003$ and $\varepsilon=0.005$, 
respectively. Meanwhile, panels (d), (e) and (f)
correspond to the case of $\phi=\pi/2$ (cos-like pulses) and for $\varepsilon=0$ (homogeneous case), 
$\varepsilon=0.003$ and $\varepsilon=0.005$, respectively.
From the panels (b), (c), (e) and (f) we can observe the strong
modifications that the nonhomogeneous character of the laser electric field
produces in the electron kinetic energy. These are related to the changes
in the electron trajectories (for details see e.g.~\cite{Yavuz12,Marcelo12A,Marcelo12OE}). In short, the electron trajectories are modified in such a way that now the electron ionizes at an earlier time and
recombines later, and in this way it spends more time in the continuum
acquiring energy from the laser electric field. Consequently, higher values
of the kinetic energy are attained. A similar behavior with the photoelectrons was observed recently in ATP using metal nanotips. According to the model presented in~\cite{Herink12} the localized fields modify the electron motion in such a way to allow sub-cycle dynamics. In our studies, however, we consider both direct and rescattered electrons (in~\cite{Herink12} only direct electrons are modeled) and the characterization of the dynamics of the photoelectrons is more complex. Nevertheless, the higher kinetic energy of the rescattered electrons is a clear consequence of the strong modifications of the laser electric field in the region where the electron dynamics takes place, as in the above mentioned case of ATP.

\subsection{ATI driven by spatially linear inhomogeneous fields: the 3D-case}

\begin{figure}[htb]
\centering
\includegraphics[width=\textwidth]{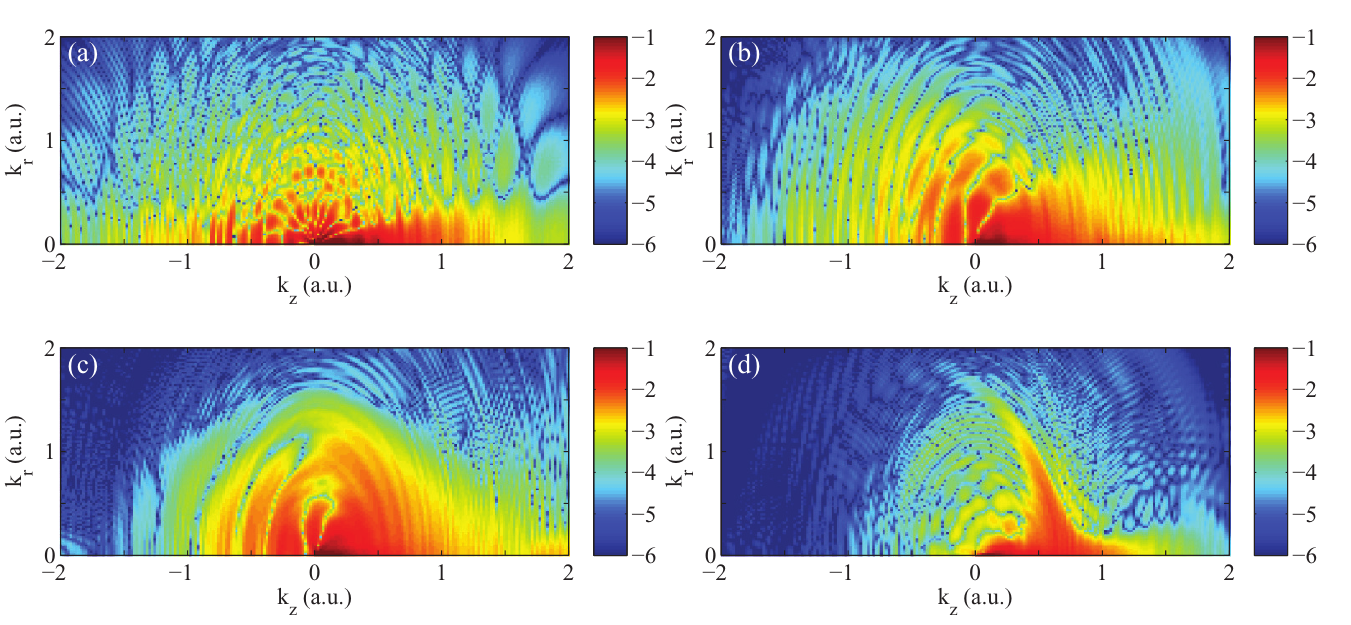}
\caption{(Color online) Two-dimensional electron momentum
distributions (logarithmic scale) in cylindrical coordinates ($k_z,k_r$)
using the exact 3D-TDSE calculation for an hydrogen atom.
The laser parameters are $I = 5.0544 \times 10^{14}$ W cm$^{-2}$ ($E_0=0.12$ a.u.) and $\lambda = 800$ nm. We have used a sin-squared shaped pulse with a total duration of four optical cycles
(10 fs) with $\phi=\pi/2$. (a) $\varepsilon = 0$
(homogeneous case), (b) $\varepsilon = 0.002$, (c) $\varepsilon = 0.003$ and (d) $\varepsilon = 0.005$.}
\label{Figure4ATI}
\end{figure}

In the following, we calculate two-dimensional electron momentum distributions for a
laser field intensity of $I=5.0544\times10^{14}$ W cm$^{-2}$
($E_0=0.12$ a.u). The results are depicted in Fig.~\ref{Figure4ATI} for
$\phi=\pi/2$. Here, panels (a), (b), (c) and (d) represent the
cases with $\varepsilon=0$ (homogeneous case), $\varepsilon=0.002$,
$\varepsilon=0.003$ and $\varepsilon=0.005$, respectively. 
By a simple inspection of Fig.~\ref{Figure4ATI} strong modifications produced by the spatial inhomogeneities in both the angular and low-energy structures can be appreciated (see~\cite{Marcelo13A} for more details).

\begin{figure}[htb]
\includegraphics[width=\textwidth]{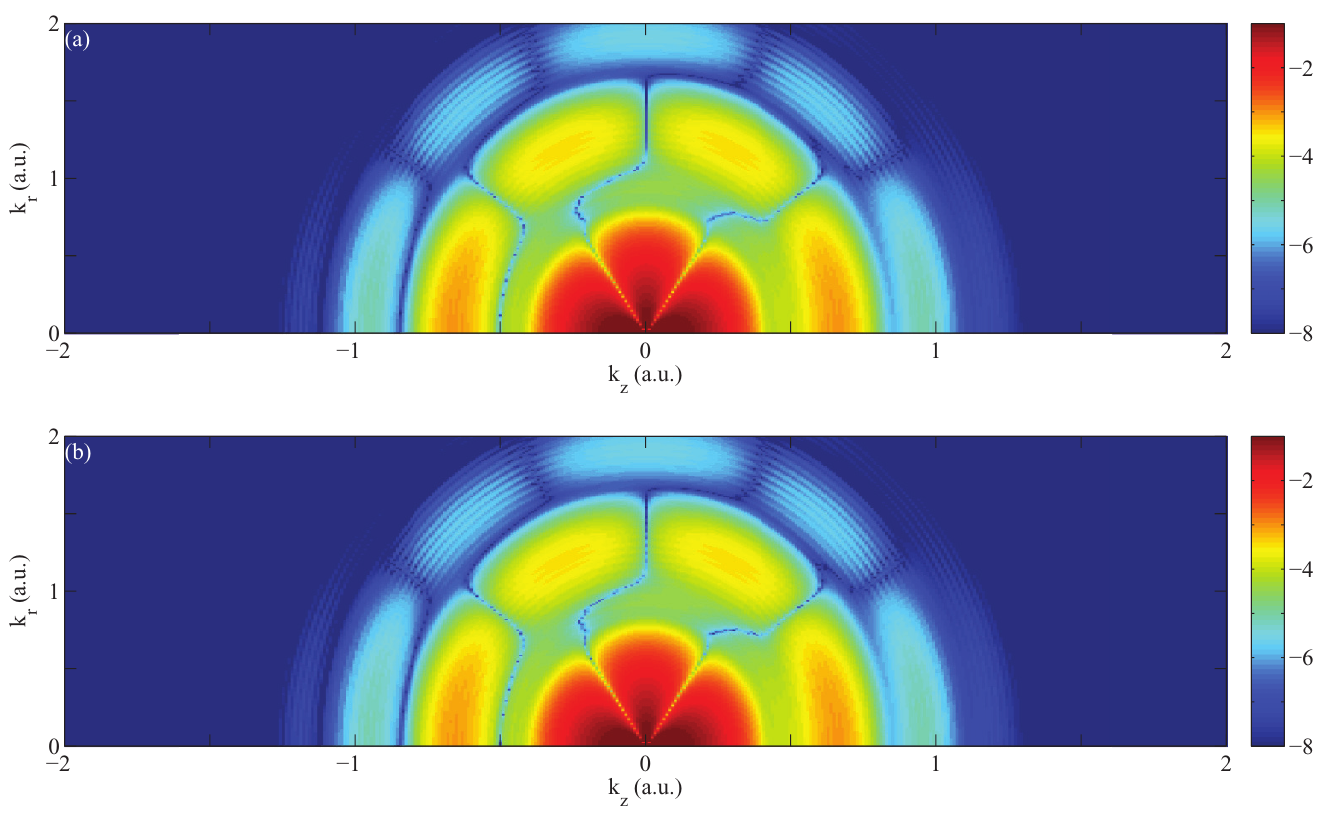}
\caption{(Color online) Two-dimensional electron momentum
distributions (logarithmic scale) in cylindrical coordinates ($k_z,k_r$)
using the exact 3D-TDSE calculation for an hydrogen atom.
The laser parameters are $E_0=0.05$ a.u. ($I=8.775\times10^{13}$ W cm$^{-2}$), $\omega_0=0.25$ a.u. ($\lambda=182.5$ nm) and $\phi=\pi/2$. We employ a laser pulse with 6 total cycles. Panel (a) corresponds to the homogeneous case ($\varepsilon=0$) and panel (b) is for $\varepsilon=0.005$.}
\label{Figure5ATI}
\end{figure}

However in the case of low intensity regime (i.e.~multiphoton regime, $\gamma\gg1$) the scenario changes radically. In order to study this regime we use a laser electric field with $E_0=0.05$ a.u. of peak amplitude ($I=8.775\times10^{13}$ W cm$^{-2}$), $\omega_0=0.25$ a.u. ($\lambda=182.5$ nm) and 6 complete optical cycles. The resulting Keldysh parameter $\gamma=5$ indicates the predominance of a multiphoton process~\cite{arbo1}. In  Fig.~\ref{Figure5ATI} we show the two-dimensional electron distributions for the two cases discussed above. For the homogeneous case our calculation is identical to the one presented in~\cite{arbo1}. We also notice the two panels present indistinguishable shape and magnitude.  Hence the differences introduced by the spatial inhomogeneity are practically imperceptible in the multiphoton ionization regime.

\subsection{Plasmonic near-fields}

In this section we put forward the plausibility to perform ATI experiments by combining plasmonic enhanced near-fields and noble gases. The proposed experiment would take advantage of the plasmonic enhanced near-fields (also known as evanescent fields), which present a strong spatial nonhomogeneous character and the flexibility to use any atom or molecule in gas phase. A similar scheme was previously presented, but now we are interested in generating highly energetic electrons, instead of coherent electromagnetic radiation. We employ 1D-TDSE by including the actual functional form of metal nanoparticles plasmonic near-fields obtained from attosecond streaking measurements. We have chosen this particular nanostructure since its actual enhanced-field is known experimentally, while for the other nanostructures, like bow-ties~\cite{Kim08}, the actual plasmonic field is unknown. For most of the plasmonic nanostructures the enhanced field is theoretically calculated using the finite element simulation, which is based on an ideal system that may deviate significantly from actual experimental conditions. For instance,~\cite{Kim08} states an intensity enhancement of 4 orders of magnitude (calculated theoretically) but the maximum harmonic measured was the 17$^{\rm{th}}$, which corresponds to an intensity enhancement of only 2 orders of magnitude (for more details see~\cite{Marcelo12OE,Marcelo12JMO}). On the other hand, our numerical tools allow a treatment of a very general set of spatial nonhomogeneous fields such as those present in the vicinity of metal nanostructures~\cite{Kim08}, dielectric nanoparticles~\cite{Zherebtsov11}, or metal nanotips~\cite{Herink12}. The kinetic energy for the electrons both direct and rescattered can be classically calculated and compared to quantum mechanical predictions (for more details see e.g~\cite{Marcelo13LPL}).

\begin{figure}[htb]
\includegraphics[width=\textwidth]{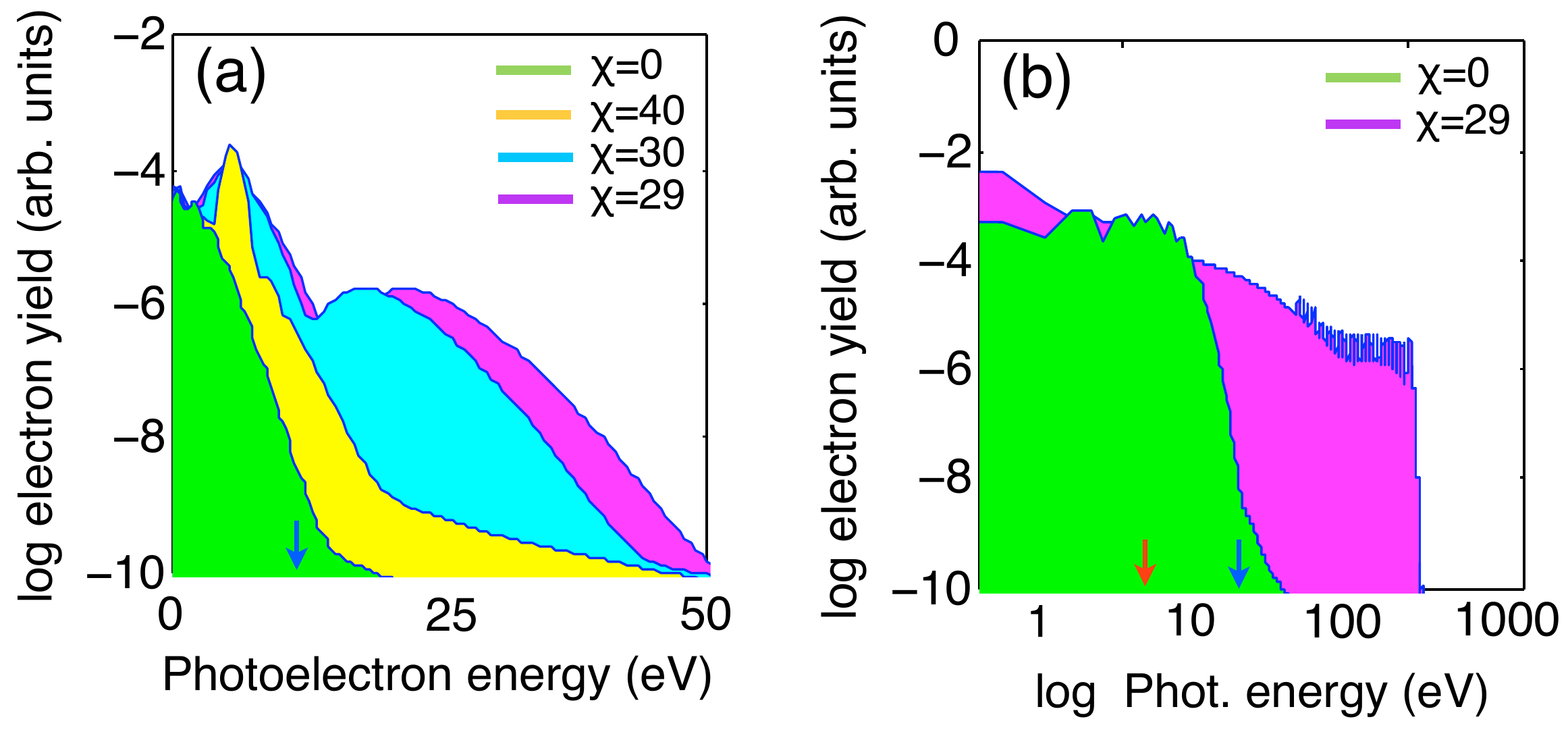}
\caption{Energy resolved photoelectron spectra for Xe atoms driven by an electric enhanced near-field. In panel (a) the laser intensity after interacting with the metal
nanoparticles is $I=2\times10^{13}$ W cm$^{-2}$. We employ $\phi=\pi/2$ (cos-like pulses) and the laser wavelength and number of cycles remain unchanged with respect to the input pulse, i.e. $\lambda=720$ nm and $n_p=5$ (13 fs in total). Panel (b) shows the output laser intensity of $I=5\times10^{13}$ W cm$^{-2}$ (everything else is the same as in panel (a)). The arrow indicates two conventional classical limits: $2 U_p$ (in red) at 5.24 eV and $10U_p$ (in blue) at 26.2 eV, respectively.}  
\label{Figure6ATI}
\end{figure}

We have employed the same parameters as the ones used in Section~\ref{HHGnearfield}, but now our aim is to compute the energy resolved photoelectron spectra. In Fig.~\ref{Figure6ATI} we present the photoelectron spectra calculated using 1D-TDSE for Xe atoms and for two different laser intensities, namely $I=2\times10^{13}$ W cm$^{-2}$ (Fig.~\ref{Figure6ATI}(a)) and $I=5\times10^{13}$ W cm$^{-2}$ (Fig.~\ref{Figure6ATI}(b)). In Fig.~\ref{Figure6ATI}(a) each curve presents different values of $\chi$: homogeneous case ($\chi\rightarrow\infty$), $\chi=40$, $\chi=35$ and $\chi=29$. For the homogeneous case there is a visible cutoff at $\approx 10.5$ eV confirming the well known ATI cutoff at $10U_{p}$, which corresponds to those electrons that once ionized return to the core and elastically rescatter. Here, $U_{p}$ is the ponderomotive potential given by $U_{p}=E_{p}^{2}/4\omega_0^{2}$. On the other hand, for this particular intensity, the cutoff at $2U_{p}$ ($\approx 2.1$ eV) developed by the direct ionized electrons is not visible in the spectrum. 

For the spatial nonhomogeneous cases the cutoff of the rescattered electron is far beyond the classical limit $10U_{p}$, depending on the $\chi$ parameter chosen. As it is depicted in Fig.~\ref{Figure6ATI}(a) the cutoff is extended as we decrease the value of $\chi$. For $\chi= 40$ the cutoff is at around $14$ eV, while for $\chi= 29$ it is around $30$ eV. The low energy region of the photoelectron spectra is sensitive to the atomic potential of the target and one needs to calculate TDSE in full dimensionality in order to model this region adequately. In this paper we are interested in the high energy region of the photoelectron spectra, which is very convenient because it is not greatly affected by the considered atom. Thus by employing 1D-TDSE the conclusions that can be taken from these highly energetic electrons are very reliable.

Figure~\ref{Figure6ATI}(b) shows the photoelectron spectra for the homogeneous case and for $\chi=29$ using a larger laser field intensity of $I=5\times10^{13}$ W cm$^{-2}$, while keeping all other laser parameters fixed. From this plot we observe that the nonhomogeneous character of the laser enhanced electric field introduces a highly nonlinear behavior. For this intensity with $\chi=29$ it is possible to obtain very energetic electrons reaching values of several hundreds of eV. This is a good indication that the nonlinear behavior of the combined system of the metallic nanoparticles and noble gas atoms could pave the way to generate keV electrons with tabletop laser sources. All the above quantum mechanical predictions can be directly confirmed by using classical simulations in the same way as for the case HHG.

Here we propose generation of high energy photoelectrons using near-enhanced fields by combining metallic nanoparticles and noble gas atoms. Near-enhanced fields present a strong spatial dependence at a nanometer scale and this behavior introduces substantial changes in the laser-matter processes. We have modified the 1D-TDSE to model the ATI phenomenon in noble gases driven by the enhanced near-fields of such nanostructure. We predict a substantial extension in the cutoff position of the energy-resolved photoelectron spectra, far beyond the conventional $10U_p$ classical limit. These new features are well reproduced by classical simulations. Our predictions would pave the way to the production of high energy photoelectrons reaching the keV regime by using a combination of metal nanoparticles and noble gases. In this kind of system each metal nanoparticle configures a laser nanosource with particular characteristics that allow not only the amplification of the input laser field, but also the modification of the laser-matter phenomena due to the strong spatial dependence of the generated coherent electromagnetic radiation.

\subsection{Emergence of a higher energy structure (HES) in ATI driven by spatially inhomogeneous laser fields}

Our final example deals with a recent study about the appearance of a higher energy structure (HES) in the energy-resolved ATI photoelectron spectra when the active media is driven by a spatially inhomogeneous laser field~\cite{LisaHES}. As was discussed throughout this contribution the theoretical approaches had not considered any spatial dependence in the field (forces) experienced by the laser-ionized electron. On the other hand, the
small spatial inhomogeneity introduced by the long-range Coulomb potential has been recently linked to a number
of important features in the photoelectron spectrum, such as Coulomb asymmetry, Coulomb focusing, and
a distinct set of low energy structures in the angle-resolved photoelectron spectra. We demonstrated that using a midinfrared laser source, with a time-varying spatial dependence in the laser electric field, such as that produced in the vicinity of a nanostructure, creates a prominent higher energy peak. This HES originates from direct electrons ionized near the peak of a single half-cycle of the laser pulse. This feature is indeed confirmed both using quantum mechanical, TDSE-based, and classical, classical trajectory monte carlo (CTMC), approaches. Interestingly, the HES is well separated from all other ionization events, with its location and energy width are strongly dependent on the properties of the spatial inhomogeneous field. As a consequence, the HES can be employed as a sensitive tool for near-field characterization in a regime where the electron's quiver amplitude is on the order to the field decay length. Additionally, the large accumulation of electrons with tuneable energy suggests a promising method for creating a localized source of electron pulses of sub-femtosecond duration using tabletop laser technology.

\section{Conclusions, Outlook and perspectives}

In this contribution we have extensively reviewed the theoretical tools to tackle strong field phenomena driven by plasmonic-enhanced fields and discussed a set of relevant results.

Nowadays, for the first time in the history of Atomic Molecular and Optical (AMO) physics we have at our disposal laser sources, which, combined with nanostructures,  generate fields that exhibit spatial variation at a nanometric scale. This is the native scale of the electron dynamics in atoms, molecules and bulk matter. Consequently, markedly and profound changes occur in systems interacting with such spatially inhomogeneous fields. Using well-known numerical techniques, based on solutions of Maxwell equations, one is able to model both the time and the spatial properties of these laser induced  plasmonic fields. This in the first  important step for the subsequent theoretical modelling of the strong-field physical processes driven by them. 

Theoretically speaking, in the recent years there has been a thoughtful and continuous activity in atto-nanophysics. Indeed,   all of the theoretical tools developed to tackle strong field processes driven by spatially homogeneous fields have beed generalized and  adapted to this new arena. Several open problems, however, still remain. For instance, the behaviour of complex systems, e.g.~multielectronic atoms and molecules, under the influence of spatial inhomogeneous fields is an unexplored area -- only few attempts to tackle this problem has been recently reported~\cite{Yavuz15,ilhan1,AlexisPRL}. In addition, and just to name another example, it was recently demonstrated that Rydberg atoms could be a plausible alternative as a driven media~\cite{rydberg}.

Diverse paths could be explored in the future. The manipulation and control of the plasmonic-enhanced fields appears as one them. From an experimental perspective this presents a tremendous challenge, considering the nanometric dimensions of the systems, although several experiments are planned in this direction, for instance combining metal nanotips and molecules in a gas phase. The possibility to tailor the electron trajectories at their natural scale is another path to be considered. By employing quantum control tools it would be possible, in principle theoretically, to drive the electron following a certain desired 'target',e.g.~a one which results with the largest possible velocity, now with a time and spatial dependent driving field. The spatial shape of this field could be, subsequently, obtained by engineering a nanostructure.

The quest for HHG from plasmonic nano-structures, joint with an explosive amount of theoretical work, begun with the controversial report of a Korean group on HHG from bow-tie metal nano structures~\cite{Kim08}. Let us mention at the end a very recent results of the same group, which clearly seems to be well justified and, as such, opens new perspectives and ways toward efficient HHG in nano-structures. In this recent article the authors demonstrate plasmonic HHG experimentally by devising a metal-sapphire nanostructure that provides a solid tip as the HHG emitter instead of gaseous atoms. The fabricated solid tips are made of monocrystalline sapphire surrounded by a gold thin-film layer, and intended to produce coherent extreme ultraviolet (XUV) harmonics by the inter- and intra-band oscillations of electrons driven by the incident laser. The metal-sapphire nanostructure enhances the incident laser field by means of surface plasmon polaritons (SPPs), triggering HHG directly from moderate femtosecond pulses of~0.1 TW cm$^{-2}$ intensities. Measured XUV spectra show odd-order harmonics up to~60 nm wavelengths without the plasma atomic lines typically seen when using gaseous atoms as the HHG emitter. This experimental outcome confirms that the plasmonic HHG approach is a promising way to realize coherent XUV sources for nano-scale near-field applications in spectroscopy, microscopy, lithography, and attosecond physics~\cite{Han2016}.  The era of the atto-nanophysics has just started. 

\section{Acknowledgments}

This work was supported by the project ELI--Extreme Light Infrastructure--phase 2 (CZ.02.1.01/0.0/0.0/15 008/0000162) from European Regional Development Fund. M. L. acknowledges the Spanish Ministry MINECO (National Plan 15 Grant: FISICATEAMO No. FIS2016-79508-P, SEVERO OCHOA No. SEV-2015-0522, FPI), European Social Fund, Fundaci\'o Cellex, Generalitat de Catalunya (AGAUR Grant No. 2017 SGR 1341 and CERCA/Program), ERC AdG OSYRIS and NOQIA, and the National Science Centre, Poland-Symfonia Grant No. 2016/20/W/ST4/00314.


%

\end{document}